\documentclass[a4paper,10pt,aps,amsfonts,amsmath,nofootinbib,showpacs,byrevtex,prd,reprint]{revtex4-1}   
\usepackage{epsfig,dcolumn}
\usepackage{graphicx}
\usepackage{color}
\usepackage{bm} 
\usepackage{amsmath}
\usepackage{amsfonts}
\usepackage{amssymb}
\usepackage{graphicx}
\usepackage{url}
\usepackage[utf8]{inputenc}

\newcommand{\be}{\begin{equation}}
\newcommand{\ee}{\end{equation}}

\definecolor{mygreen}{rgb}{0.0,0.4,0.0}
\definecolor{myred}{rgb}{1.0,0.0,0.0}
\def\lsim{\mathrel{\rlap{\lower4pt\hbox{\hskip1pt$\sim$}}
    \raise1pt\hbox{$<$}}}
\def\gsim{\mathrel{\rlap{\lower4pt\hbox{\hskip1pt$\sim$}}
    \raise1pt\hbox{$>$}}}

\def \be {\begin{equation}}
\def \ee {\end{equation}}
\newcommand{\ba} {\begin{align}}
\newcommand{\ea}{\end{align}}
\newcommand{\beqa}{\begin{eqnarray}}
\newcommand{\eeqa}{\end{eqnarray}}

\begin{document}

\title{Neutral Pion Photoproduction in a Regge Model} 

\author{V. Mathieu}\email{mathieuv@indiana.edu}
\affiliation{Center for Exploration of Energy and Matter, Indiana University, Bloomington, IN 47403}
\affiliation{Physics Department, Indiana University, Bloomington, IN 47405}
\author{G. Fox}
\affiliation{School of Informatics and Computing, Indiana University, Bloomington, IN 47405, USA}
\author{A. P. Szczepaniak}
\affiliation{Center for Exploration of Energy and Matter, Indiana University, Bloomington, IN 47403}
\affiliation{Physics Department, Indiana University, Bloomington, IN 47405}
\affiliation{Thomas Jefferson National Accelerator Facility, 
Newport News, VA 23606, USA}

\date{\today}

\begin{abstract}
The reaction $\gamma p \to \pi^0 p$ is investigated in the energy range above the resonance region. The amplitudes include the leading Regge singularities in the cross-channel and correctly describe the differential cross section for beam energies above 4 GeV and for the  $s-$channel scattering angle $\cos\theta_s\ge 0.6$. The energy dependence of the beam asymmetry and the reaction  $\gamma n \to \pi^0 n$ seem  is quantitative consistent with the Regge-pole dominance. 
 \end{abstract}

\pacs{13.75.Gx, 13.85.Fb, 11.55.Jy, 12.40.Nn}

\maketitle

\section{Introduction}\label{intro} 
Single pion photoproduction on a nucleon is one of the key reactions in hadron physics. At low energies, it is  used to excite nucleon resonances while at high energies it can be used to test  
 predictions of Regge theory, {\it e.g.} factorization of Regge poles~\cite{Irving:1977ea}. The two regimes are analytically connected and relations {\it e.g.} finite energy sum rules (FESR), can be derived to constrain resonance parameters by the cross-channel Reggeons~\cite{Worden:1972dc}.  In recent years the CLAS experiment at JLab collected data on  single pion production using photon beams with energies ranging from $E_\gamma = 1.275\mbox{ GeV}$ to $ 5.425\mbox{ GeV}$~\cite{CLAS}.  This energy range overlaps with both, the resonance and the Regge regions.  Once the data is analyzed it will open the possibility to perform finite energy sum rules studies of pion photoproduction based on data from a single experiment. This will reduce systematic uncertainties and possibly help to clarify some of the outstanding theoretical issues encountered in earlier studies~\cite{Odorico:1976wn}. Single pion photoproduction will also be among the first reactions studied with the newly completed GlueX detector that will use highest energy photons from the recently upgraded CEBAF accelerator. 

There are several neutral pion photoproduction models developed to describe the nucleon resonance region~\cite{Drechsel:2007if,Workman:2011vb,Anisovich:2012ct,Kamano:2013iva}, which is not the case of the high-energy regime. Regge description has not been updated in the recent past, with the exception of Ref.~\cite{Sibirtsev:2009kw}. In view of the forthcoming measurements it is therefore necessary to revisit the theoretical models. 
 This the main focus of this paper. According to Regge theory, at asymptotically large center of mass energies reaction amplitudes are determined by the right-most singularly in the complex angular momentum plane of the cross-channel partial waves. Except for elastic scattering, which is dominated by the Pomeron, these are the Regge poles. A single Regge pole contribution factorizes into a product of  ``couplings/residues" and ``propagators". This property enables classification of Regge poles similar to that of elementary particles~\cite{Gribov:2009zz}. As the center of mass energy decreases, sub-leading angular momentum plane singularities become relevant including Regge poles with lower intercept   {\it aka.} daughter trajectories and Regge cuts. 
  Since the ultimate goal is to connect the Regge and resonance regions and make predictions for the energy range of the CLAS measurement, in this paper we also explore these sub-leading contributions. 
 We focus on scattering in the forward direction where 
    the dominant Regge singularities originate from the  $t-$channel exchanges. 
  The goal is to find a simple, albeit consistent with Regge theory,  parametrization that once the CLAS data is analyzed, can be applied in a simultaneous study of the resonance and Regge regions {\it e.g.} using  finite energy sum rules. 
   
The paper is organized as follows. 
In Section~\ref{sec:formalism} we discuss the $t$-channel amplitudes and 
   their connection to the direct, $s$-channel observables. Conventions and details of calculations are given in Appendix \ref{app:t} and \ref{app:s}.   Specification of the Regge exchange model is given in Section \ref{sec:model}. Analysis of existing data is summarized in Section \ref{sec:results} and 
     predictions for the CLAS energy range and for $E_\gamma = 9\mbox{ GeV}$, relevant for the upcoming experiments at JLab are given in Section \ref{sec:pred}. Conclusions are summarized in Section~\ref{sec:concl}.

\section{Formalism}\label{sec:formalism}
The reaction $\gamma(k,\lambda) N(p_2) \to \pi(q) N(p_4)$ describing pion with momentum $q$, the photon with momentum $k$ and helicity $\lambda$, and a pair of nucleons with momenta $p_2$ and $p_4$ is given in terms of four invariant amplitudes,  which are linearly related to 
 four helicity amplitudes.  The scalar amplitudes are functions of two independent Mandelstam variables, $s=(k+p_2)^2$ and $t=(k-q)^2$. 
The helicity amplitudes are frame dependent. The two relevant frames for our discussion are the $s-$channel and $t-$channel frames. They correspond to the center of mass frame of the reaction $\gamma N \to \pi^0 N$ for the $s-$channel and $\gamma \pi^0 \to \bar N N$ for the $t-$channel, respectively. Helicity amplitudes in the $s-$channel are used to compute observables. At high energies and small scattering angles the $s$-channel amplitudes are dominated  by singularities in the complex angular plane of the $t-$channel. Helicity amplitudes in the $t-$channel are therefore needed to identify the  allowed Regge exchanges.  Detailed analysis of the $t-$channel helicity amplitudes and their quantum numbers is given in Appendix~\ref{app:t}. The $s-$channel helicity amplitudes and the observables are discussed in Appendix~\ref{app:s}. In this section we summarize the main results. 
  
The invariant amplitudes, $A_i(s,t)$ multiply four independent tensors constructed from the photon polarization vector, two Dirac spinors and particle momenta constrained to fulfill global 
 symmetry requirements and gauge invariance. The tensors  are conventionally chosen 
  from the Chew-Goldberger-Low-Nambu (CGLN) basis~\cite{Chew:1957tf}, and the helicity amplitudes are given by 
  \begin{subequations} \label{eq:defT}
\begin{align}
A^s_{\mu_4,\mu_2\mu_1}(s,t) &= \bar u_{\mu_4}(p_4) \sum_{i=1}^4 A_i(s,t) M_i\  u_{\mu_2}(p_2), \\
A^t_{\lambda_4\lambda_2,\lambda_1}(s.t) &= \bar u_{\lambda_4}(p_4) \sum_{i=1}^4 A_i(s,t) M_i v_{\lambda_2}(-p_2),
\end{align}
\end{subequations}
 with the subscripts $s$ and $t$  referring to the $s$ and $t$ channels,  and in the following we use $\mu_i$ and $\lambda_i$ to denote particle helicities in the two channels. 
  The tensors  $M_i$ are given in \eqref{eq:defM}.  The scalar functions $A_i$ 
 have dynamical singularities in $s$ and $t$ while 
the helicity amplitudes have additional singularities arising from the kinematical factors in 
Eq.~\eqref{eq:defT}.  To identify $t$-channel reggeons it is necessary to identify $t$-channel helicity amplitudes free from kinematical singularities.  Kinematical singularities are related to presence of spin and can be related to singularities of the Wigner-$d$ functions. 
 The rotational  functions can be written as 
 $d_{\lambda'\lambda}^J(z_t)=\xi_{\lambda'\lambda}(z_t)P_{\lambda'\lambda}^J(z_t)$ where $P^J$ is a polynomial 
and
\be \label{eq:xi}
\xi_{\lambda'\lambda}(z_t) = \left(\frac{1-z_t}{2}\right)^{\frac{1}{2} |\lambda'-\lambda| } \left(\frac{1+z_t}{2}\right)^{\frac{1}{2} |\lambda'+\lambda| },
\ee
with $ \lambda=\lambda_1-\lambda_3=\lambda_1$, $\lambda'=\lambda_2-\lambda_4$.  
 As shown in Appendix~\ref{app:t} singularities in $s$ of the $t$-channel helicity amplitudes can be removed by dividing helicity  amplitudes by $\xi$, 
\begin{align} 
\widehat T_{\lambda'\lambda} = \xi^{-1}_{\lambda'\lambda}(z_t)
 A^t_{\lambda_4\lambda_2,\lambda_1}(s,t).  
\end{align}
The remaining singularities of $\widehat T$  in $s$ are dynamical in nature. 
Since Reggeons have well defined quantum numbers in the $t-$channel, they will contribute to specific linear combination of the invariant amplitudes $A_i$.  Helicity amplitudes 
 are not eigenstates of parity and the $t-$channel parity conserving helicity amplitudes (PCHAs) correspond to linear combination, 
\cite{collins, spearman}
\begin{align}
\widehat T^\eta_{\lambda'\lambda} = \frac{1}{\sqrt{2}}\left( \widehat T_{\lambda'\lambda} +\eta(-1)^{\lambda'}\ \widehat T_{\lambda'-\lambda} \right).
\end{align}
where $\eta$ is  the $t$-channel naturality, {\it i.e.}  $\eta = P(-)^J$ where $P$ is the intrinsic parity and $J$ is the spin of the exchange Reggeon in the $t-$channel. 
As shown in  the Appendix~\ref{app:t}, the relations between PCHAs and invariant amplitudes are~\cite{salin}
\begin{subequations}
\begin{align}
\widehat T^+_{01} &=  - 2 k_t \sqrt{t} (-A_1+2M A_4) \\
\widehat T^-_{01} &= -4 p_t k_t(A_1+t A_2) \\
\widehat T^+_{11} &=  -2 k_t (2MA_1-t A_4) \\
\widehat T^-_{11} &=  \phantom{+}4 p_t k_t \sqrt{t} A_3
\end{align}
\end{subequations}
with  $M$ being  the  nucleon mass. The photon momentum, $k_t$ and the proton momentum $p_t$ evaluated in the  $t-$channel {\it cf.} Eq.~\eqref{eq:kint}, contain the remaining kinematical singularities in $t$. We can now define four amplitudes, free of kinematical 
singularities that have well defined quantum numbers ($\eta$ and $CP$) in the $t-$channel. These are, 
   \begin{subequations}\label{eq:Finv}
\begin{align}
F_1 &= -A_1+2MA_4, & \eta&=+1, & CP&=+1,\\
F_2&= A_1+t A_2 , & \eta&=-1, & CP&=-1,\\
F_3&= 2M A_1-t A_4, & \eta&=+1, & CP&=+1,\\
F_4 &=A_3,  & \eta&=-1, & CP&=+1.
\end{align}
\end{subequations}
Only negative charge conjugation, $C=-1$ exchanges, couple to $\gamma\pi^0$. 
  For positive naturality, these correspond to vector trajectories $\omega$ ($I^G=0^-$) and $\rho$ ($I^G=1^+$) and contribute to $F_1$ and $F_3$.  For negative $\eta$ the axial-vector trajectories 
   $h$ ($I^G=0^-$) and $b$ ($I^G=1^+$) contribute to $F_2$. There are no known mesons contributing to $F_4$. The lowest  mesons in the $t-$channel contributing to $F_4$ would be the $\rho_2$ and $\omega_2$ with $J^{PC}=2^{--}$.  Therefore in the following we ignore the  $F_4$ (although there are some indications that $F_4$ might not be exactly zero \cite{Barker:1974vm}). 

Using the invariant amplitudes $F_i$ defined in Eq.~\eqref{eq:Finv} one can compute all observables. 
 In particular we are interested in the differential cross section and the single polarization asymmetries. The beam asymmetry is $\Sigma=(d\sigma_\bot-d\sigma_\parallel)/(d\sigma_\bot+d\sigma_\parallel)$ where $d\sigma_\perp$ ($d\sigma_\parallel$) is the differential cross section with photon polarization along the $x$ ($y$) axis and the $z$ axis along the direction of the photon momentum and $y$ perpendicular to the reaction plane. The target (recoil) asymmetry is defined as $T (R)= (d\sigma_\uparrow-d\sigma_\downarrow)/(d\sigma_\uparrow+d\sigma_\downarrow)$ and measures the asymmetry of the spin polarization of the
target (recoil) nucleon.

At high energies, keeping only the leading $s$ dependence in the kinematical factors relating the differential cross section to the scattering amplitude,  one finds~\cite{Barker:1974vm}\footnote{We corrected the sign of $F_1$ in the target and recoil asymmetries, {\it cf.}  Appendix \ref{app:s}.}
\begin{subequations}\label{eq:CS}
\begin{align}
 \frac{d\sigma}{dt} &\approx  \frac{1}{32\pi} \left[\frac{|F_3|^2- t|F_1|^2}{4M^2- t} + |F_2|^2 - t |F_4|^2 \right] \\
\Sigma \frac{d\sigma}{dt} &\approx  \frac{1}{32\pi} \left[\frac{|F_3|^2- t|F_1|^2}{4M^2- t} - |F_2|^2 + t |F_4|^2 \right] \\
T \frac{d\sigma}{dt} &\approx  \frac{1}{16\pi} \sqrt{- t}\ \text{Im}\,\left[\frac{F_3 F_1^*}{4M^2- t} + F_4F_2^* \right]\\
R \frac{d\sigma}{dt} &\approx  \frac{1}{16\pi} \sqrt{- t}\ \text{Im}\,\left[\frac{F_3 F_1^*}{4M^2- t} - F_4F_2^* \right].
\end{align}
\end{subequations}
The differential cross section in physical units ($\mu\text{b GeV}^{-2}$) is obtained by multiplying the right hand sides by the conversion factor $1 =389.4\ \mu\text{b GeV}^{2}$. 
These asymptotic formulas are useful for identifying contributions from the individual Regge contributions to the amplitudes $F_i$.  In numerical calculations that follow, for the differential cross section, we use the complete expression
\begin{align}\nonumber
 \frac{d\sigma}{dt} =  \frac{1}{64\pi} &\frac{|k_t|^2}{4 M^2 E_\gamma^2} \bigg[
\phantom{+} 2|\sin\theta_t|^2 \left( |2 p_t F_2|^2 - t |F_1|^2 \right) \\ \nonumber
& +  (1-\cos\theta_t)^2 \left| F_3-2\sqrt{t} p_t F_4\right|^2 \\ 
& +  (1+\cos\theta_t)^2 \left| F_3+2\sqrt{t} p_t F_4\right|^2  \bigg],
\label{eq:CSfull} 
\end{align}
where $E_\gamma$ is the beam energy in the laboratory frame.  To check the validity of the asymptotic approximation, in  Section~\ref{sec:pred} we compare the results obtained using Eq.~\eqref{eq:CS} and Eq.~\eqref{eq:CSfull}. In calculations of the spin asymmetries, however, we use the asymptotic  formulas of Eq.~\eqref{eq:CS} since the finite-$s$ corrections cancel in the ratio of cross sections. 

Most of the available data comes from the proton target with only a few measurements of the 
 differential cross sections on the neutron. The corresponding amplitudes are related by isospin symmetry. In the $t$-channel, the isospin decomposition for each of the four invariant amplitudes is  
  is  given by \cite{Chew:1957tf}
\be
A^a_{\alpha\beta} = A^{(+)}\delta^{a3}\delta_{\alpha\beta} + A^{(-)}\frac{1}{2}[\tau^a,\tau^3]_{\alpha\beta} + A^{(0)} \tau^a_{\alpha\beta}.
\ee
with $A^{(+)}$, $A^{(-)}$ and $A^{(0)} $ corresponding to $t$-channel isospin $I^G=0^-$, $1^-$ and  $1^+$  and  $a$, $\alpha$, $\beta$ being the isospin indices for the pion, the two nucleons, respectively. Only $A^{(+)}$ and $A^{(0)}$ contribute to $\pi^0$ photoproduction.
We note that the isovector exchange contribution contributes with opposite sign to proton and neutron amplitudes, {\it i.e.} 
\begin{subequations} \label{eq:isospin}
\begin{align}
A(\gamma p \to \pi^0 p) = A^{(+)}+A^{(0)},\\
A(\gamma n \to \pi^0 n)  = A^{(+)}-A^{(0)}.
\end{align}
\end{subequations}


\section{The Regge Model} \label{sec:model}
In this section we specify the model for the $t$-channel kinematical singularity free amplitudes 
  $F_i(s,t)$. The contribution of a Regge pole, $R(s,t)$ and a Regge-Pomeron cut $R_c(s,t)$ to $F_i(s,t)$ have asymptotic energy dependence determined by the Regge trajectory $\alpha(t)$\cite{collins,Collins:1971ff,Irving:1977ea}. 
  The residues are  analytical in  $t$ in the $s$-channel physical region and have zeros that are 
   forced by spin considerations. In particular, in the physical region of the $t$-channel, net helicity in either,  the  $\gamma \pi^0$ or the $N\bar N$ vertex cannot exceed $J=\alpha(t)$ for non-negative,  integer values of $J$. In addition, in the $s$-channel physical region 
    the amplitude cannot have singularities other than threshold branch points. A simple model that builds in these constraints is given by, 
    
\begin{align} \label{eq:Regge}
R(s,t) & = \frac{\pi}{\Gamma(\alpha(t))} \frac{1-e^{-i\pi\alpha(t)}}{2\sin \pi \alpha(t)} \left(\frac{s}{s_0}\right)^{\alpha(t)-1}, \\
R_c(s,t) & = \frac{1}{\log(s/s_0)} \frac{\pi}{\Gamma(\alpha_c(t))} \frac{1-e^{-i\pi\alpha_c(t)}}{2\sin \pi \alpha_c(t)} \left(\frac{s}{s_0}\right)^{\alpha_c(t)-1}.
\end{align}
The energy dependence yields the expectation $s^2d\sigma/dt\propto s^{2\alpha(t)}$ once the extra energy power coming from the half-angle factor in Eq.~\eqref{eq:xi} is included. 
These expressions can in principle be multiplied by an analytical function of $t$, that can be, for example  fixed by the data. To minimize the number of parameters in the model, we will attempt to fit the data with minimal such modifications (see below) observing that the scale $s_0$ already leads to an  
  exponential fall off with $t$ for $s > s_0 = 1\mbox{ GeV}^2$ which is typical.  
For positive values of spin, $J=\alpha>0$, the signature factor $1-e^{-i\pi\alpha}$ is finite for the right  exchanges,  {\it i.e.} odd-spin mesons lying on the vector or axial-vector trajectories and it vanishes, canceling  the zero of $\sin\pi\alpha$  for the wrong spin exchanges {\it i.e.} even-spin mesons. 
For negative odd-values of $\alpha$ there would be unphysical poles. The $\Gamma$ function in the denominator is introduced to remove these poles from the  $s$-channel physical region. 
It also  leads to zeros of the amplitude at non-positive even integer values of $\alpha$. 
As discussed above Eq.~\eqref{eq:Regge}  $\alpha=0$ is indeed forbidden by spin considerations since it is less than the magnitude of net helicity in the  $\gamma\pi^0$ vertex. 
In principle there is no need for the amplitude to vanish at negative even-integer values of $\alpha$. 
Thus the particular choice of using the $\Gamma$ function to cancel the unphysical poles of the so-called Regge propagator,  $1/\sin\pi\alpha$,  has to be confronted with the data and can in principle be modified if needed. 
We will further explore the consequences of this choice in Section \ref{sec:results} and \ref{sec:pred}. 


For values of $\alpha$ which are not too far from positive integers Regge pole amplitudes are 
 similar to this of particle exchange. We therefore use a one-particle-exchange model for a 
  vector and axial-vector meson to  impose further constraints. 
  The $t$-channel amplitude for a vector meson exchange, {\it e.g.} $V=\rho,\omega$ is given by ({\it cf.} Appendix~\ref{app:t})
\begin{align} \nonumber
\frac{i \varepsilon_{\alpha\beta\mu\nu} k^\mu\epsilon^\nu q^\alpha}{t-m^2_V} \bar u_4
\left[g^V_4 \gamma^\beta + g^V_1 \gamma^{[\beta}\gamma^{\sigma]}(p_2-p_4)_\sigma\right] v_2 \\
 =  \bar u_4 \left[ \frac{ g^V_4 M_4 + g^V_1 (tM_1-M_2)}{ t-m^2_V}\right] v_2,
\label{eq:PTvector}
\end{align}
 where $ \bar u_4 \equiv \bar u(p_4,\mu_4)$ and $v_2 \equiv v(-p_2,\mu_2)$ with the right hand side  written in terms of the CGLN basis tensors $M_i$. Similarly for the axial vectors,  $A=b,h$ one finds, 
\begin{align}\label{eq:PTaxial}
g^A_2 \epsilon^\mu
\left[k\cdot q g_{\mu\alpha} - q_\mu k_\alpha \right] \frac{ \bar u_4 \gamma_5 p^\alpha  v_2}{t-m^2_A}  & = 
\frac{\bar u_4 g^A_2 M_2  v_2}{t-m^2_A}.
\end{align}
Comparing with Eq.~\eqref{eq:defT} on concludes that,  for the vector contribution, $A_1\propto g_1 t, A_2\propto -g_1$ and $A_4\propto g_4$. We also note that in the $s$-channel frame $g_4$ corresponds to helicity non-flip coupling and $g_1$ to helicity flip coupling. Near the pole, the  tree-level propagator $1/(t-m^2_V)$, 
  corresponds to $1/\sin\pi\alpha$.  We also note that, as expected,  the vector exchanges contribute only to $F_1$ and $F_3$. Also, as expected we find that the axial vector change contributes only to 
  $F_2$ via $A_2 \propto g^A_2$.


The data, as it will be described in the next section, indicate that corrections to the Regge pole approximation are needed. We then anticipate and add in our model the cut associated with the vector trajectory. Their role will be clarified by comparing to the data.  The $t-$dependence of Regge-Pomeron  cut residues is, {\it a priori}, different from that of the Regge poles. For simplicity, however,  we will use the same $t-$dependence as for the corresponding pole. 
Taking these considerations into account leads to the following expressions for the invariant amplitudes $F_i$ 
\begin{align} \nonumber
F_1 &= (-g_1 t+2Mg_4) R^V(s,t)  + (-g^c_1 t+2Mg^c_4) R_c(s,t) ,\\ \nonumber
F_2  &=  g_2 t R^A(s,t),\\ \nonumber
F_3&= (2M g_1 - g_4) t R^V(s,t) + (2M g^c_1 - g^c_4) t R_c(s,t),\\
F_4 &=0.
\label{eq:FiRegge}
\end{align}
We do not include cut contributions in the axial exchange amplitudes. 
In Eqs \eqref{eq:FiRegge}, ($g_1$, $g_2$, $g_4$, $g^c_1$, $g^c_4$) are real parameters to be determined by fitting the data.  Because of the connection with the one-particle exchange model 
 they can be related to the products of $\gamma \pi^0$ and $NN$ couplings of vectors and axial vectors~\cite{Yu:2011zu}. Since matching between particle exchange and Regge amplitudes is only exact at the $t$-channel pole, the  polynomial $t$ dependence in Eqs~\eqref{eq:FiRegge} is rather arbitrary, we keep it nevertheless to allow for more flexibility in describing the $t$-dependence. 

The subscripts $V$ and $A$ in Eq.~\eqref{eq:FiRegge} specify the Regge trajectory $\alpha = \alpha_V(t)$ or $\alpha = \alpha_A(t)$ to be used in the Regge amplitudes of Eq.~\eqref{eq:Regge}. 
We assume exchange degeneracy {\it i.e.} the same trajectory $\alpha_V$ for $\omega$ and $\rho$, and the same axial trajectory, $\alpha_A$ for  $b$ and $h$ which, including the cut, we take to be linear,  
\begin{align} \nonumber
\alpha_V(t) &= \alpha_{V0}+ \alpha_V' t, &
\alpha_A(t) &= \alpha_{A0}+ \alpha_A' t, \\
\alpha_c(t) &= \alpha_{c0}+ \alpha_c' t.
\end{align}

Furthermore, from Eq.~\eqref{eq:isospin} it follows that for the  $g_1=\pm g_1^\rho+g_1^\omega$,  $g_4=\pm g_4^\rho+g_4^\omega$ and  $g_2=\pm g_2^b+g_2^h$ with the upper/lower sign referring the proton/neutron amplitudes, respectively. 
In addition to the five couplings we need to determine the six trajectory parameters $(\alpha_{V0},\alpha_{V}',\alpha_{A0},\alpha_{A}',\alpha_{c0},\alpha_{c}')$. Trajectory parameters are constrained by the meson mass-spin relations and are expected to be approximately given by~\cite{Irving:1977ea} 
\begin{subequations}\label{eq:OSalpha}
\begin{align}
\alpha_V(t) &= 1+ 0.9(t-m^2_\rho) & & \sim 0.5  + 0.9 t \\
\alpha_A(t) &= 1+ 0.7(t-m^2_{b_1}) & &\sim 0.7 t
\end{align}
\end{subequations}
In a simple model for Pomeron exchange~\cite{Irving:1977ea}, the Regge-Pomeron 
 cut trajectory has an intercept, $\alpha_{c0} = \alpha_{V0}$  and a slope close to that of the Pomeron, {\it i.e} we expect the cut trajectory to approximately give by  
  \begin{align}
\alpha_c(t) & \sim 0.5  + 0.2 t.
\end{align}

\section{Results}\label{sec:results}
\subsection{Data selection and interpretation} \label{sec:data}
\begin{figure*}[htb!]
	\includegraphics[width=0.49\linewidth]{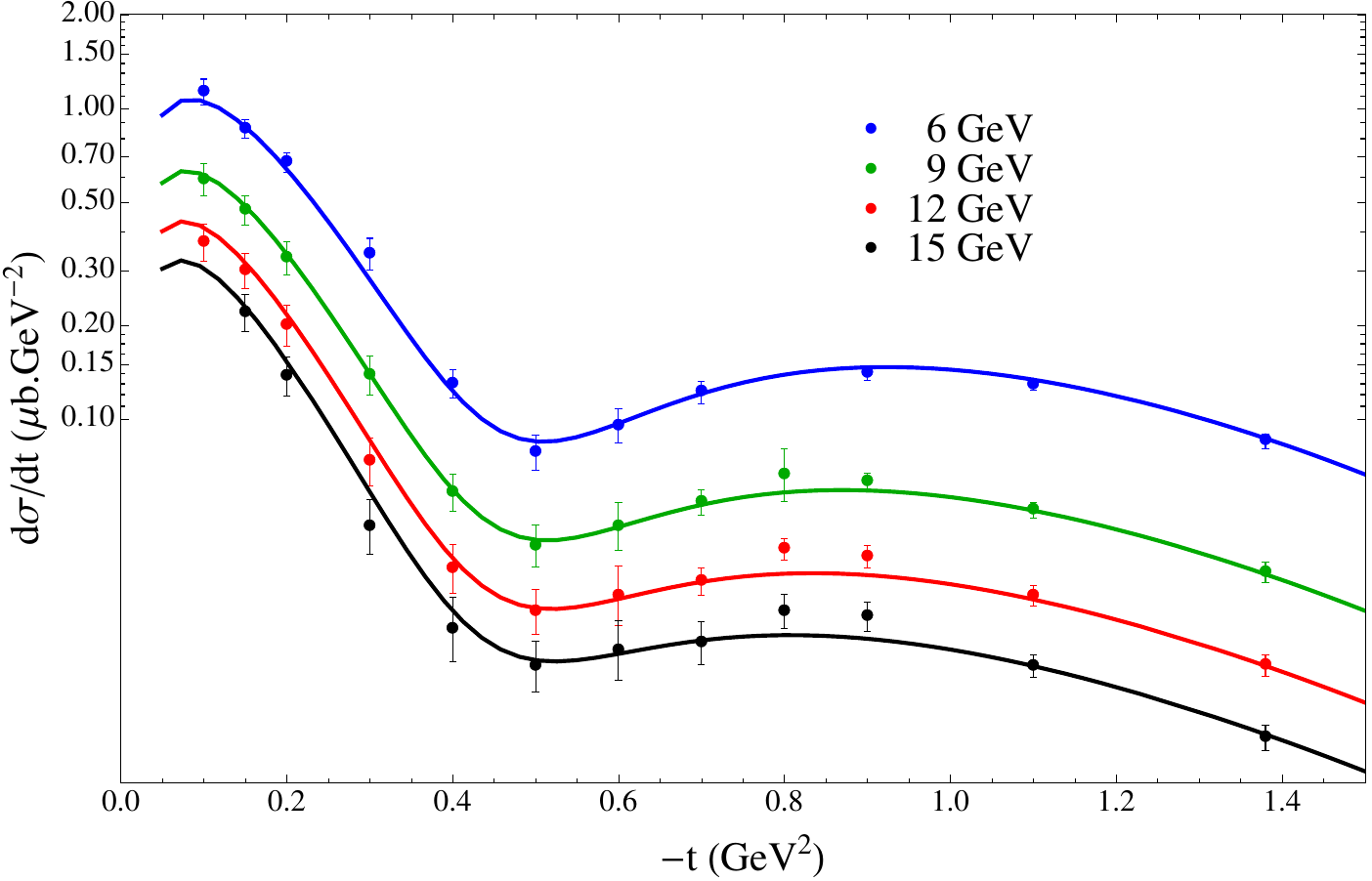} 
	\includegraphics[width=0.49\linewidth]{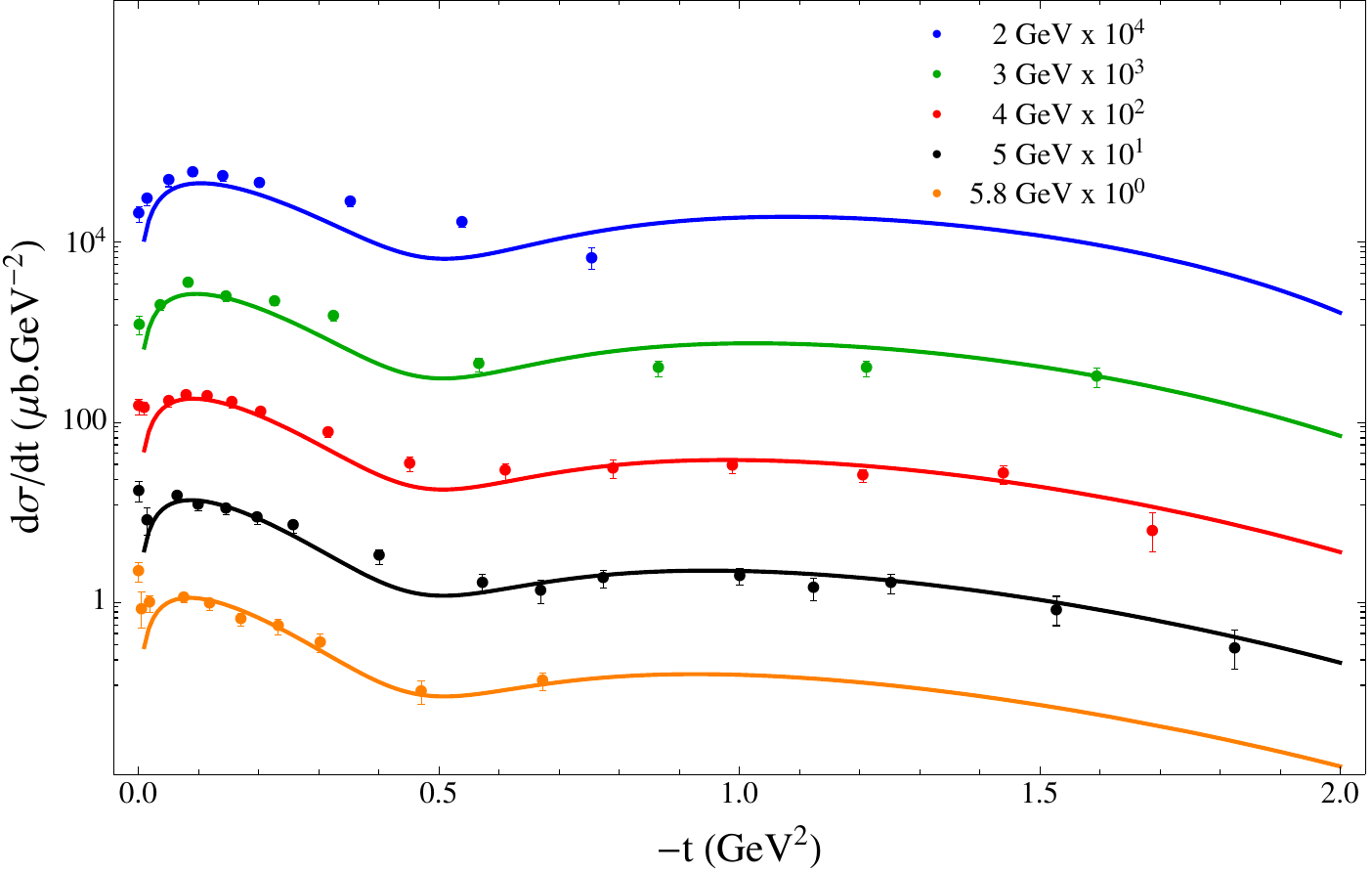}
	\includegraphics[width=0.49\linewidth]{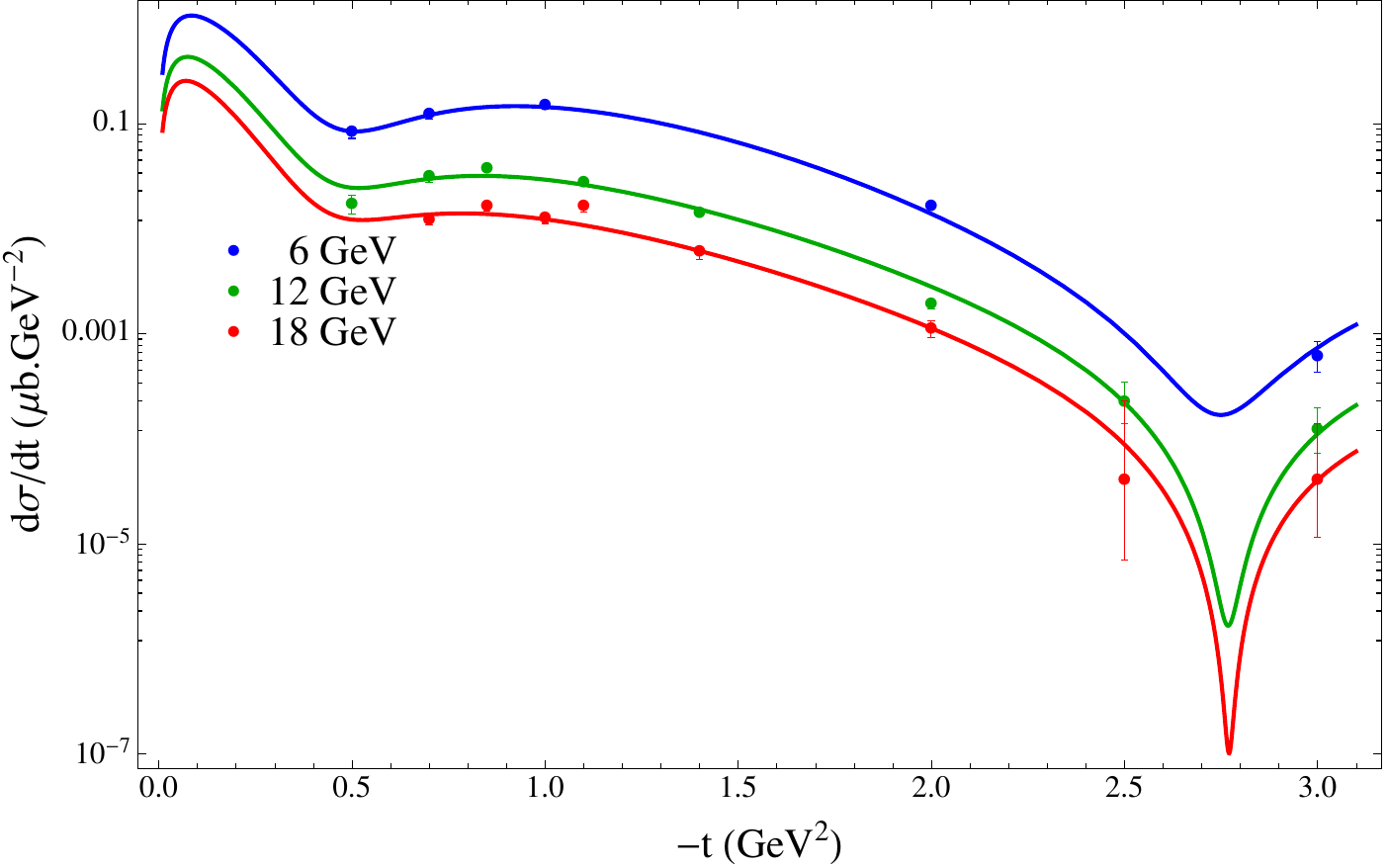} 
	\includegraphics[width=0.49\linewidth]{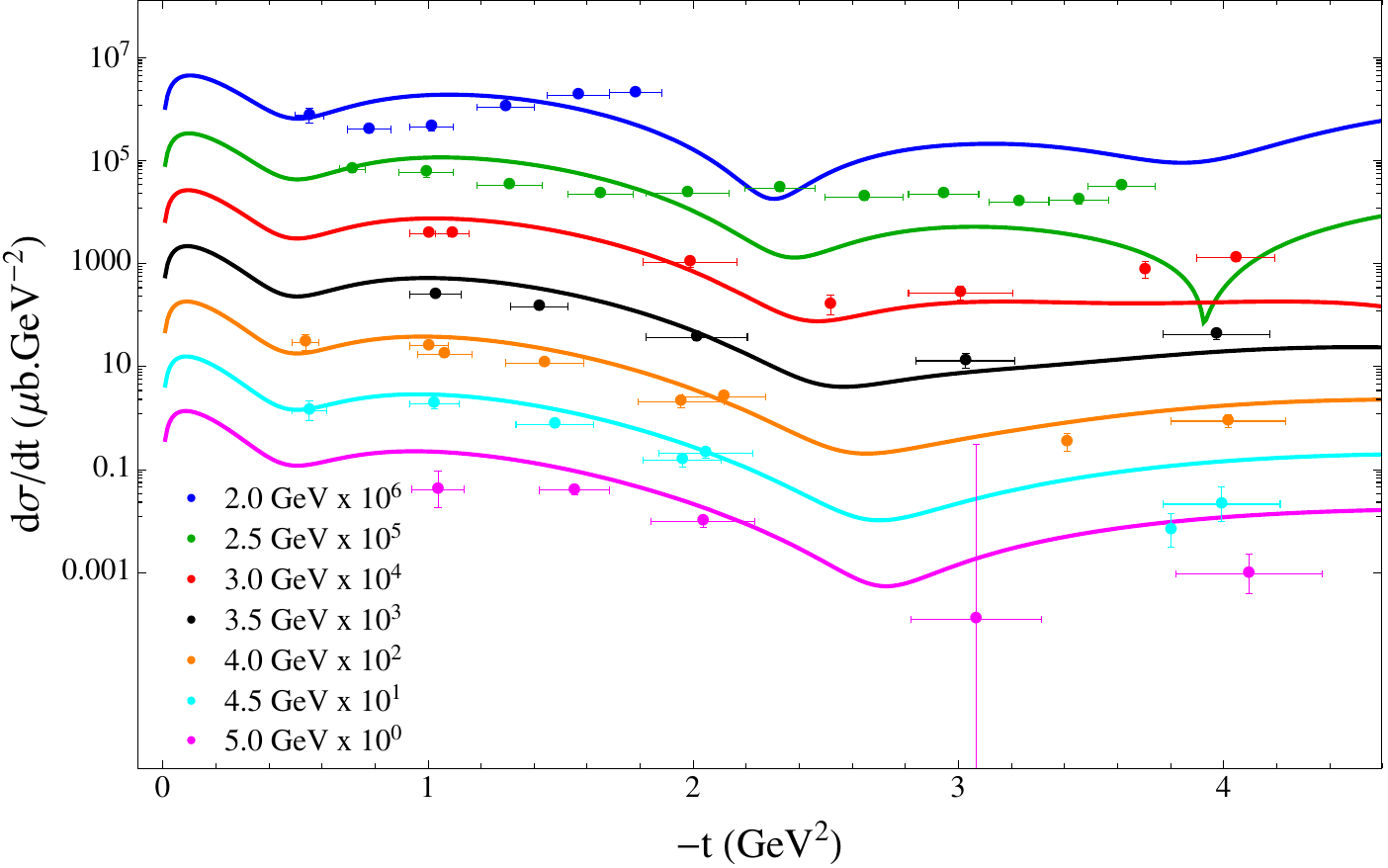}
        \caption {(Color online). Differential cross section. Only the data in the region $E_\gamma\ge 4$ GeV and $\cos\theta_s\ge0.6$ are used in the fit (see text). The model results in this region of are  represented by solid lines. The dashed lines represent the extrapolation outside the fitting region.  Data from \cite{Anderson:1971xh, Braunschweig:1973vu, Bolon:1967zy, Barish:1974qg,Anderson:1976ph}.} \label{fig:dataCS} 
\end{figure*}

\begin{figure*}[htb!]
 	\includegraphics[width=0.49\linewidth]{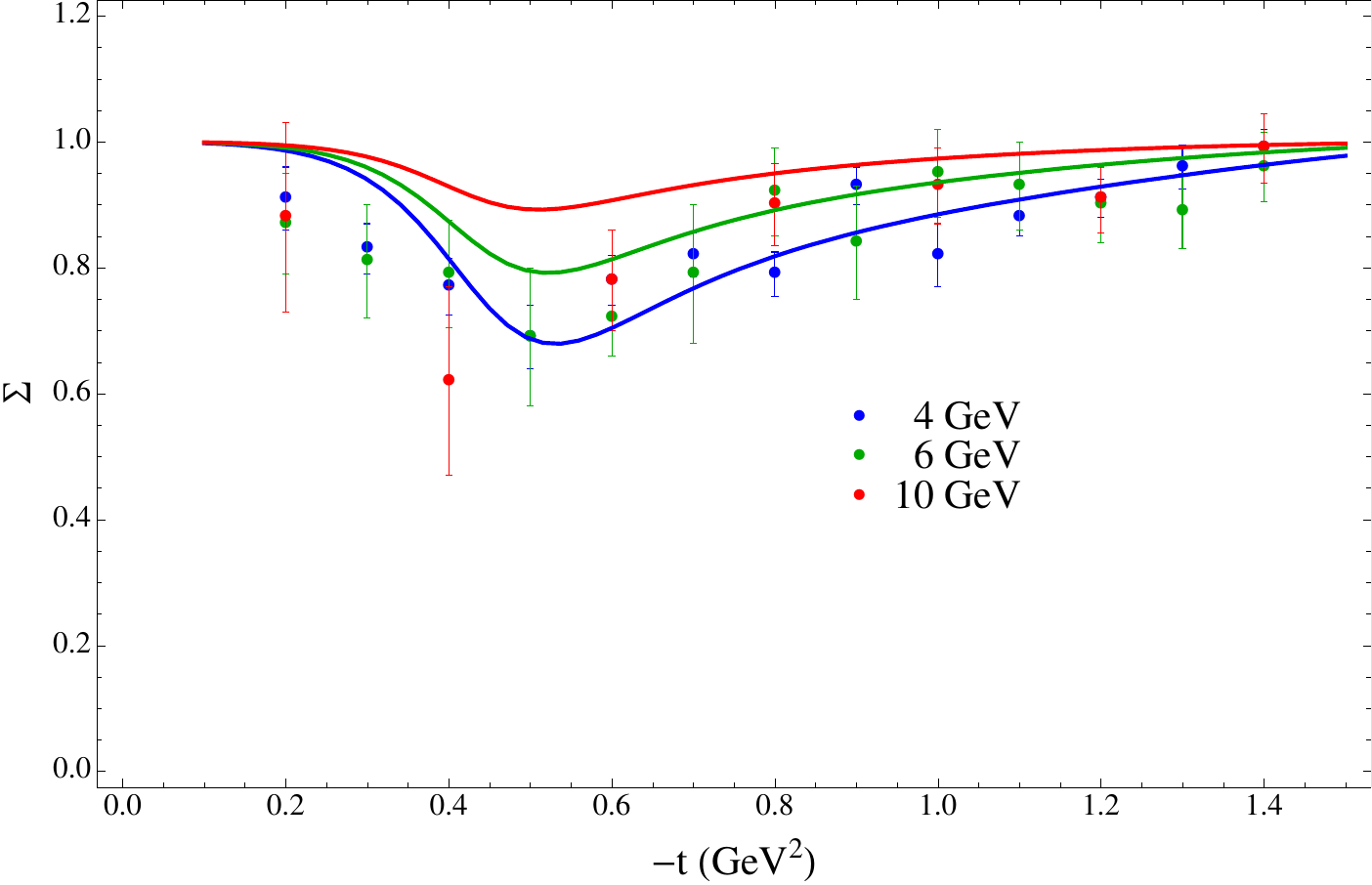}
         \includegraphics[width=0.49\linewidth]{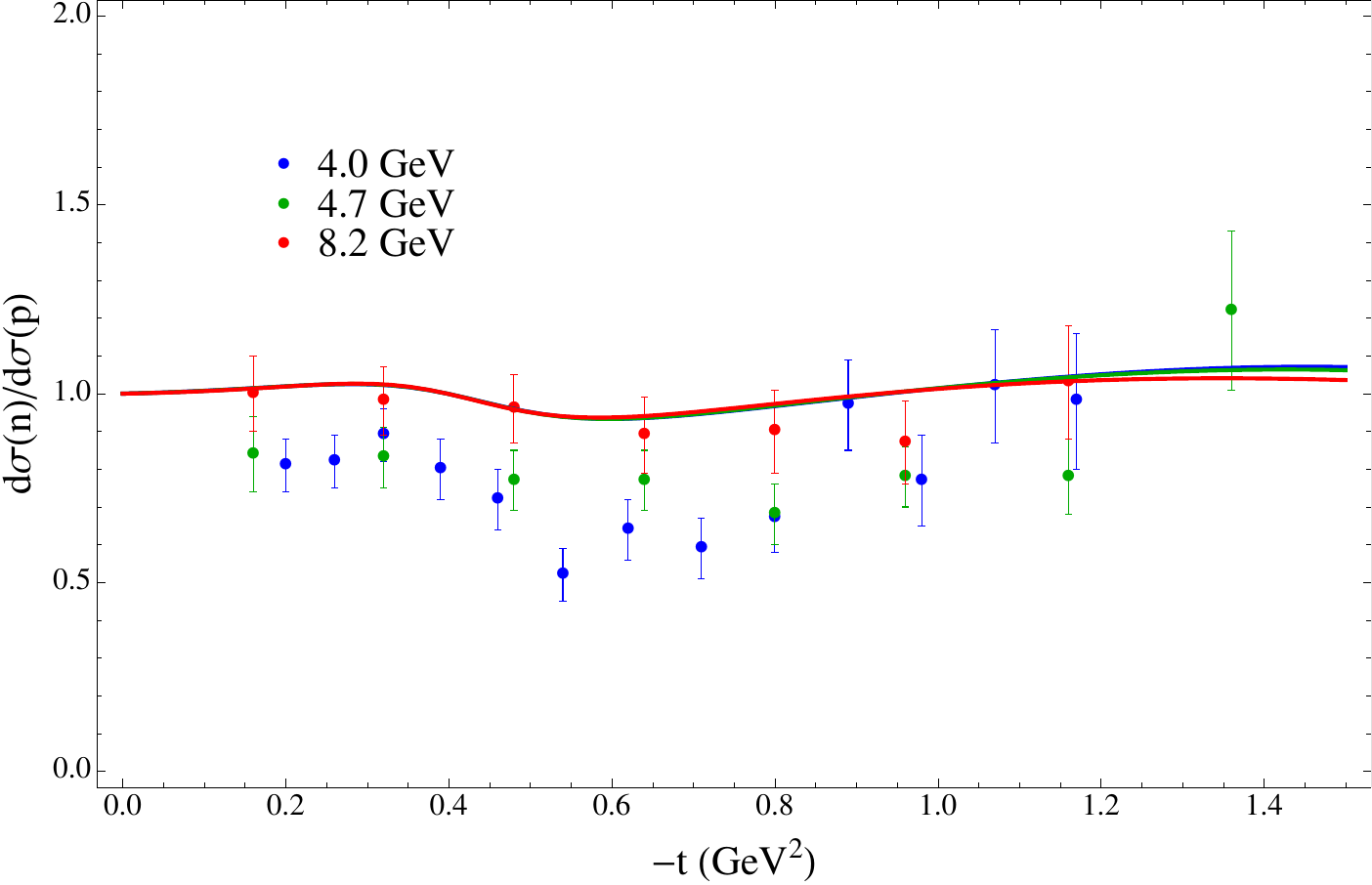}
        \caption {(Color online). Beam asymetry (left) and ratio of differential cross sections with neutron target to proton target (right). Data from \cite{Anderson:1971xh} (left) and \cite{Osborne:1973ed,Braunschweig:1973wd} (right)}
        \label{fig:dataBA}
\end{figure*}

The $t-$channel exchanges govern in the physics of the $s-$channel region only for large energies and small angles. The exact region of validity of Regge theory will be deduced by analyzing the data. In the region $E_\gamma \ge 2 $ GeV and $|t|< 3$ GeV$^2$ there are data on differential cross sections~\cite{Anderson:1971xh,Barish:1974qg,Bolon:1967zy,Braunschweig:1973vu,Anderson:1976ph} shown in Fig.~\ref{fig:dataCS}, the beam asymmetry ~\cite{Anderson:1971xh}, shown in Fig.~\ref{fig:dataBA}, the ratio of differential cross section on neutron to proton targets \cite{Osborne:1973ed,Braunschweig:1973wd}, shown in Fig.~\ref{fig:dataBA}, and on target and recoil asymmetries \cite{Booth:1972qp,Bienlein:1973pt, Deutsch:1973tg}, shown in Fig.~\ref{fig:dataTA}.

The differential cross section in the energy range $E_\gamma=6-15$ GeV has a dip at 
 $t\sim -0.5$ GeV$^2$. This value of momentum transfer is close to the wrong signature point of the vector trajectory {\it i.e.} $\alpha_V(t)=0$. As explained in the previous section, vector exchange is expected to vanish at $\alpha_V=0$ since it corresponds to a nonsense point, {\it i.e.} unphysical helicity coupling.  The minimum seen in the data has therefore a clear interpretation within the Regge theory, however, a single Regge pole model would imply vanishing of the cross section at the non-sense point which is inconsistent with the data. The disappearance of the dip at lower energies, $E_\gamma \lsim 2$ GeV can be used as signal for inapplicability of the simple Regge model. The data in the intermediate region $E_\gamma=2-6$ GeV, however, does not give a precise determination of the energy where the simple Regge picture breaks down.  
 
The beam asymmetry is sensitive to exchange on trajectories corresponding to mesons with 
 negative naturality, since it is given by 
 \begin{equation} 
  \Sigma=\frac{|\omega+\rho|^2-|h+b|^2}{|\omega+\rho|^2+|h+b|^2}
  \end{equation}
where each term corresponds to a single Regge amplitude. 
 With the vector contribution being close to zero at $t\sim -0.5$ GeV$^2$, the beam asymmetry 
  is predated to be $\Sigma \sim -1$. In the other limit, with the axial vector being close to zero,
    $\Sigma=+1$ and deviations from this value measure the strength of the axial vector 
    Regge trajectory contributions, as seen in Fig.~\ref{fig:dataBA}. For not too large momentum transfers, $|t| < 2.5$ GeV$^2$, that value of $\alpha$ for the  vector trajectory is larger than for the axial, $\alpha_V(t)>\alpha_A(t)$, {\it cf.} Eq.~\eqref{eq:OSalpha}, thus as  energy increases, the  contribution of axial exchanges relative to vector exchanges decreases. The beam asymmetry is therefore expected to approach one as the energy increases. The data in Fig.~\ref{fig:dataBA}, however, shows an approximatively constant beam asymmetry as a function of energy.  

The ratio of the differential cross sections on the neutron compared to the proton target also indicates presence of exchanges other than single Regge poles. This ratio is given by   
\begin{equation} 
\frac{d\sigma (n)/dt}{d\sigma (p)/dt} =  \frac{|\omega-\rho|^2+|h-b|^2}{|\omega+\rho|^2+|h+b|^2}.
\end{equation} 
Since, phenomenologically it is observed that the isoscalar exchanges $\omega$ and $h$ have 
  trajectory slopes approximatively equal to those of their vector partners, $\rho$ and $b$, 
   the ratio is expected to be relatively energy independent. The data shown in the right panel 
   in Fig. \ref{fig:dataBA} indicates a significant deviations from this expectation. 

We conclude that  qualitative features of the data are consistent with the single Regge pole approximations, however, quantitative description requires inclusion of other contributions, 
{\it e.g.} daughter trajectories or cuts. 

\subsection{Cut versus daughter}
\label{cvd} 
As discussed above, near the nonsense point  $\alpha=0$ ($t\sim -0.5$ GeV$^2$),  differential 
 cross section is small but non-vanishing. Zero in the vector Regge exchange amplitude 
 can be lifted by either axial Regge poles or corrections to the single Regge pole model. In order to determine which contribution dominates over the vector exchange we compare energy dependence of the differential cross section at $t=-0.5$ GeV$^2$ to that at $t=-0.1$~GeV$^2$, where the vector pole is expected to dominate.  For the axial or daughter trajectory this ratio would decrease with energy since both have an intercept which is  smaller than that of the vector trajectory. On the other hand, if it is a Regge-Pomeron cut the ratio would, up to logarithmic corrections,  be approximatively energy independent since  the  intercept of the cut  is similar to that of the pole, {\it cf.} Eq.~\eqref{eq:Regge}. 
 The measured ratio between differential cross sections at these two values of momentum transfer 
  is  $6.9\%$, $6.6\%$ and $6.5\%$ at $E_\gamma=6$, 9 and 12 GeV, respectively. This 
   is almost energy independent and we conclude that cuts might be more relevant than sub-leading  Regge poles in filling up the zero at $t\sim -0.5$ GeV$^2$. 
Another discriminator between  cuts and poles is the $t$-dependence. 
 The logarithmic slope shown in Fig.~\ref{fig:dataCS}, in the region $t\in[-0.1,-0.4]$ GeV$^2$ where the pole dominates is larger than that in the region $t\in[-0.9,-1.4]$ GeV$^2$. Since cut has a smaller 
  slope than a pole we conclude that at the larger value of $|t|$ where the leading pole is suppressed it is  the cut that dominated the differential cross section. 
   To illustrate the difference between the  vector-Regge-Pomeron cut model and a model with a 
   sub-leading pole  we compare their predictions for the differential cross section. 
   In $\pi^0$ photoproduction, as shown later, the dominant Regge-pole contribution comes from  the $\omega$ exchange, which is predominantly helicity-non-flip at the nucleon vertex 
   in the $s$-channel. We thus place the $\omega$ contribution into the $A_4$ amplitude
    and take, 
     \begin{subequations}
\begin{align}
F_1 & = 2M g_4^p R(\alpha_V) + 2M g_4^c R(\alpha_c) \\
F_3 & =-t g_4^p R(\alpha_V) -t g_4^c R(\alpha_c) 
\end{align}
\end{subequations}
with $g_4^p = 1$ and $g_4^c=0.1$, representing a $10\%$ contribution of the $\omega$-Pomeron cut at the amplitude level. For a model with a daughter trajectory, we take 
\begin{subequations}
\begin{align}
F_1 & = 2M g_4^p R(\alpha_V) + 2M g_4^d R(\alpha_d) \\
F_3 & =-t g_4^p R(\alpha_V) -t g_4^d R(\alpha_d) 
\end{align}
\end{subequations}
with $g_4^p = 1$ and $g_4^d=0.5$. The couplings were chosen so that both models yield 
 comparable cross sections. 

\begin{figure}[htb!]
	\includegraphics[width=\linewidth]{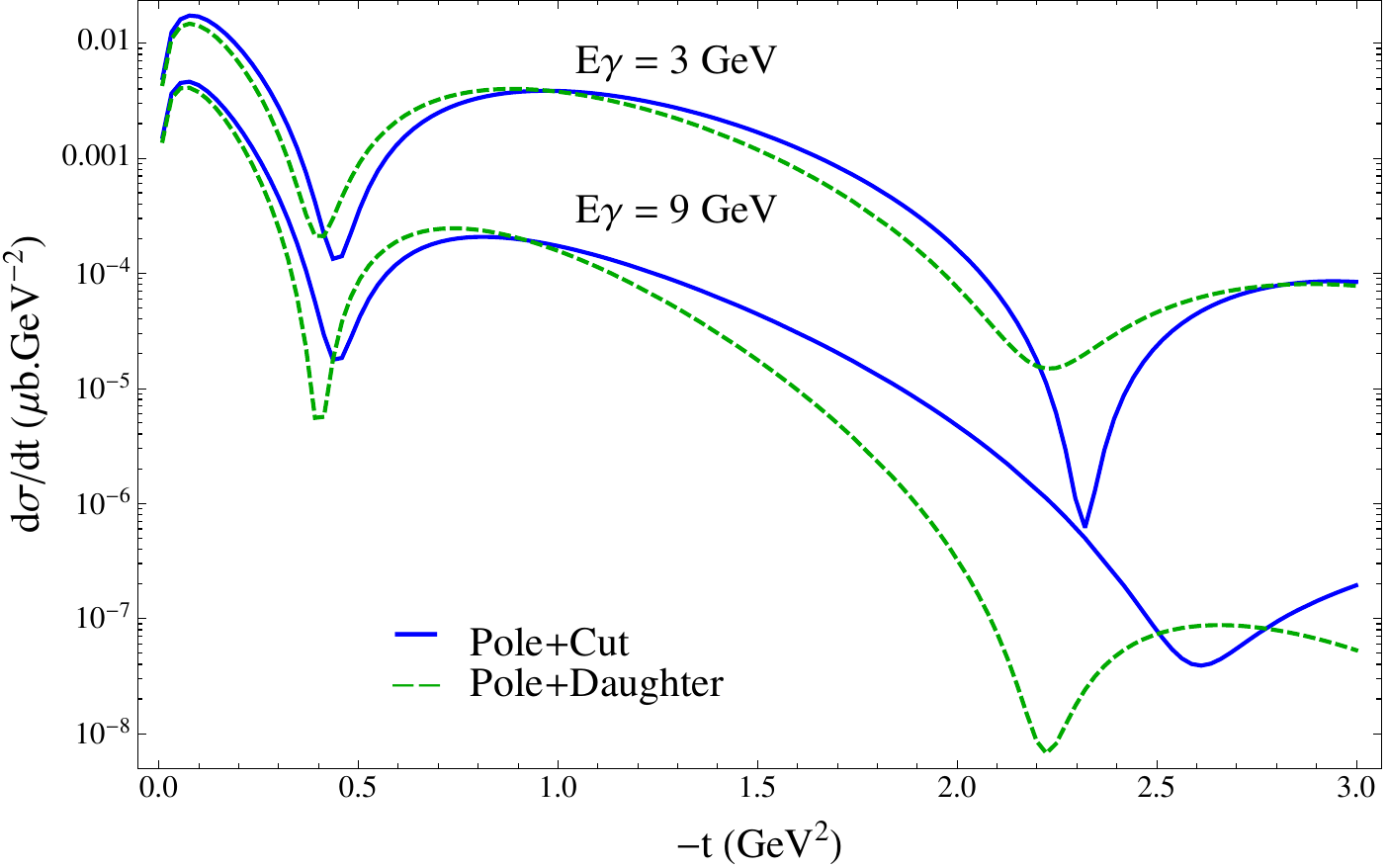}
        \caption {(Color online). The two simple models discussed in the text for a beam energy of 3 and 9 GeV.} 
        \label{fig:cutdau}
\end{figure}
The predicted deferential cross section for the two models at two photon energies $E_\gamma=3$ and 9 GeV is shown in Fig.~\ref{fig:cutdau}.  As expected, the  logarithmic slope in the range $1 \mbox{ GeV}^2 <|t| <2 \mbox{ GeV}^2 $ is different in the two models, the cut resulting in a smaller slope and dominating in this region of momentum transfer. Both models, have the first minimum coinciding with zero of the $\omega$ Regge pole but the energy dependence of the zero is deferent. Specifically, the ratio between the first maximum, at around $t=-0.1$ GeV$^2$ and the first minimum, at around $t=-0.4$ GeV$^2$ is almost energy independent for the model with the cut. The weak energy dependence of the dip and the smaller logarithmic slope at larger momentum  transfers of the model with the cut make it phenomenologically more appealing compared to the model with an additional pole. 

We also note, that in the model with the cut,  the position of the second dip, around $|t|\sim 2.2-2.6$ GeV$^2$ seems to move with energy more than in the model with a second pole. 
With the existing data it is not possible to test this prediction, however, it might be possible with the forthcoming CLAS data \cite{CLAS}.

\subsection{Fitting procedure} \label{sec:fit}
The parameters of the Regge model from Sec.~\ref{sec:model} are determined  as follows. 
 First we fix the vector Regge pole and the associated Regge cut parameters by fitting 
 the differential cross section of photoproduction on the proton target.
 This does not determine the isospin of the vector exchange. 
   The axial vector pole contribution to the differential cross section is small, ({\it cf.} Sec. ~\ref{cvd}) and in the fit to the cross section it is ignored. The axial vector Regge pole parameters are determined by fitting the beam asymmetry, once the vector exchanges are fixed by the differential cross section.   The target and recoil asymmetries are not included  in the fits and constitute a prediction of the model. 
  
We include in the fit the data from~\cite{Anderson:1971xh} and~\cite{Barish:1974qg} in the kinematical range $E_\gamma \ge 6$ GeV and for $\cos\theta_s\ge0.6$. We also include data from~\cite{Braunschweig:1973wd} and \cite{Anderson:1976ph}  for $E_\gamma \ge 4$ GeV and $\cos\theta_s\ge0.6$. We exclude the very forward region  $|t|< 0.01\mbox{ GeV}^2$ since it is dominated by the Primakoff process\footnote{For a parametrization of the Primakoff effect in neutral pion photoproduction see Refs~\cite{Sibirtsev:2009kw} and~\cite{Laget:2005be}.}.
We do not include the data from~\cite{Bolon:1967zy}, which is compatible with the other data,  but has much wider $t$-bins and no data points close to the forward direction. We have found that the lower  energy data $E_\gamma<4$ GeV results in a significantly larger $\chi^2/\text{d.o.f.}$ compared to the other sets. Therefore we fix model parameters using data above 4 GeV and predict the cross section in the lower energy range $E_\gamma<4$ GeV. For $E_\gamma\ge 4$ GeV the model reproduces the data in the whole range of $|t|\le 3$~GeV$^2$.  Overall the fit results in $\chi^2/\text{d.o.f.}=3.43$ and for the parameters of the vector pole and cut are find (with $\alpha'$ is in GeV$^{-2}$)
\begin{subequations} \label{eq:fit1}
\begin{align}
g_1 &= \phantom{+} 1.24\pm 1.56 \text{ GeV}^{-4},  & 
\alpha_{V0} & = 0.54 \pm 0.03\\
g_4 &= -6.68\pm 0.80 \text{ GeV}^{-3},  & 
\alpha_{V}' & = 1.34 \pm 0.08\\
g_1^c &=-2.36 \pm 0.36 \text{ GeV}^{-4},  & 
\alpha_{c0} & = 0.43 \pm 0.03\\
g_4^c &=-4.26 \pm 0.99 \text{ GeV}^{-4}, &
\alpha^\prime_c &= 0.16\pm0.01.
\end{align}
\end{subequations}
Although we did not constrain parameters of the trajectories, the fit finds the vector trajectory 
 consistent with expectations, {\it cf.} Eq.~\eqref{eq:OSalpha}.  

With the vector Regge pole and cut parameters determined using the high-energy data, in 
Fig.~\ref{fig:dataCS} we compare model prediction with the data in the lower-energy region,  $E_\gamma\ge 2$~GeV. The model (solid lines) is extrapolated outside the fitting region ({\it i.e.} outside $E_\gamma\ge4$ GeV and $\cos\theta_s\ge0.6$) (dashed lines). It appears that the simple Regge pole plus a cut is qualitatively consistent with the data outside this region up to $t=-3$ GeV$^2$, although the data in this region is rather sparse and it is impossible to clearly identify the region of applicability of the Regge theory.

As discussed in Sec.~\ref{sec:data},  we assume that the main contribution to beam asymmetry comes form the axial vector Regge poles. From a fit to the beam asymmetry in the energy range $4\le E_\gamma \text{(GeV)}\le 10$ with the vector Regge exchanges fixed by Eq.~\eqref{eq:fit1} we find
\begin{subequations}
\begin{align}
g_2 &= - 9.74\pm 2.96 \text{ GeV}^{-4},  \\
\alpha_{A0} & = -0.22 \pm 0.33\\
\alpha^\prime_A &= \phantom{-} 1.08\pm 0.21 \text{ GeV}^{-2}.
\end{align}
\end{subequations}
with   $\chi^2/d.o.f=1.78$.  The large uncertainty obtained for the intercept is not surprising. It originates 
 from the discrepancy in the energy dependence between the data and the model with a single axial pole {\it cf.} Sec.~\ref{sec:data}.

Since the differential cross section and the beam asymmetry do not discriminate between isovector and isoscalar Regge poles the coupling parameters in Eq.~\eqref{eq:fit1} are the sum of the two exchanges. The helicity flip and non-flip pole residues in the $s-$channel are proportional to $g_1$ and $g_4$ respectively. If we assume that  in the $s-$channel the $\omega$ trajectory is dominantly helicity non-flip  so that, $g_4=g^\omega_4$, the $\rho$ trajectory is helicity flip,  {\it i.e.} $g_1=g^\rho_1$ \cite{Irving:1977ea},  and we neglect the $h$ trajectory, {\it i.e.} $g_2=g_2^b$, we can make  a prediction for the ratio of the differential cross section on the 
 neutron and proton targets. The two are related by a sign change in  $g_1$ and $g_2$ {\it cf.} Eq.~\eqref{eq:isospin}.  A better agreement with the data is found if we consider the cut couplings to be both induced by a Pomeron-$\omega$ exchange, {\it i.e.} we do not flip the sign of $g_1^c$ and $g_4^c$ for a neutron target. We compare this prediction with the data ~\cite{Osborne:1973ed, Braunschweig:1973wd} in the energy range $4\le E_\gamma \text{(GeV)}\le 8.2$ in Fig.~\ref{fig:dataBA}. The angular 
 distribution of the data is well reproduced by our theoretical prediction. However, as we already commented in Sec.~\ref{sec:data}, the energy dependence of our model is only qualitatively consistent with the data.  The degeneracy between $\omega$ and $\rho$ trajectories produced a ratio of the differential cross section on the neutron and proton targets independent of the energy. 
  
\begin{figure*}[htb!]
	\includegraphics[width=0.49\linewidth]{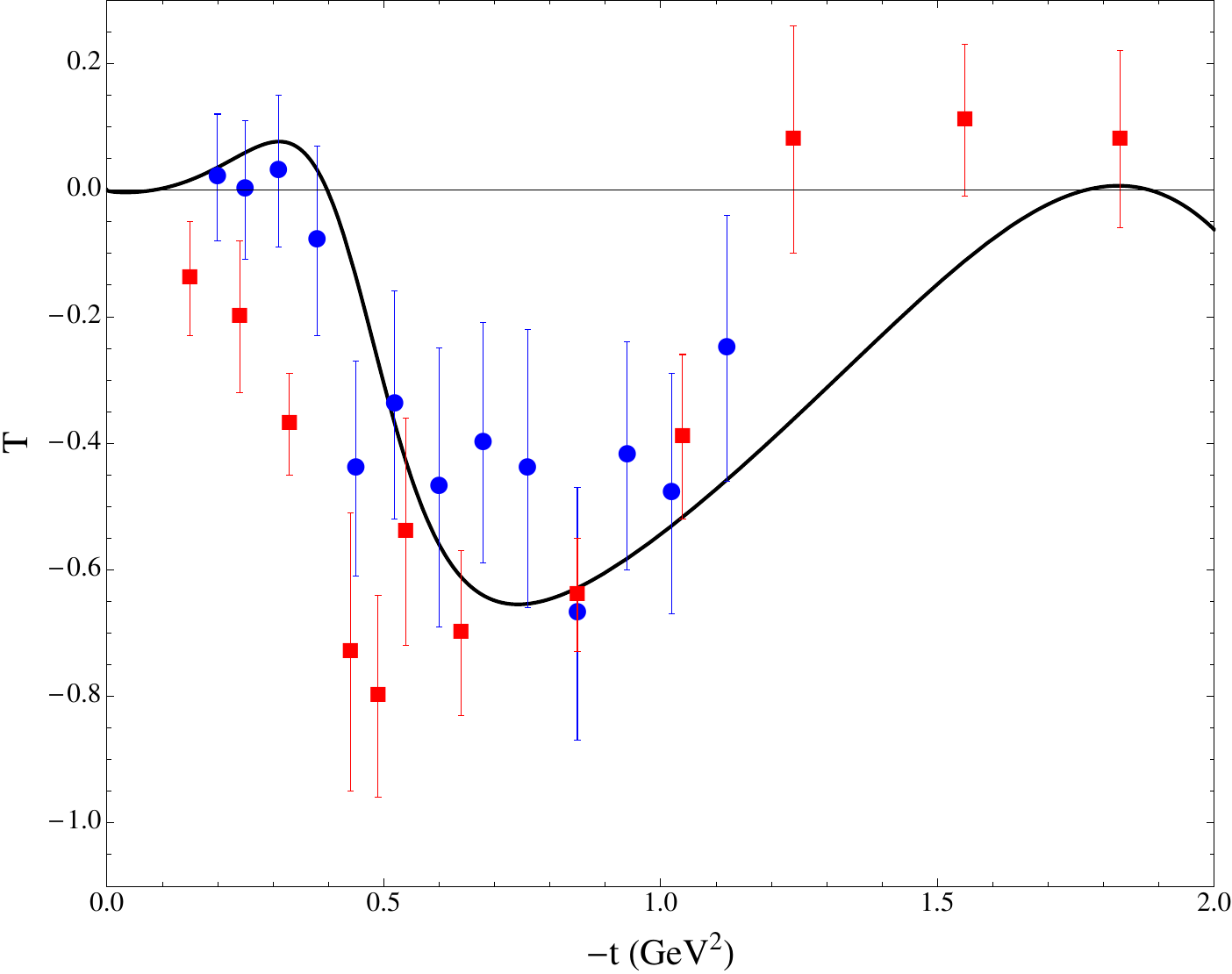}
	\includegraphics[width=0.49\linewidth]{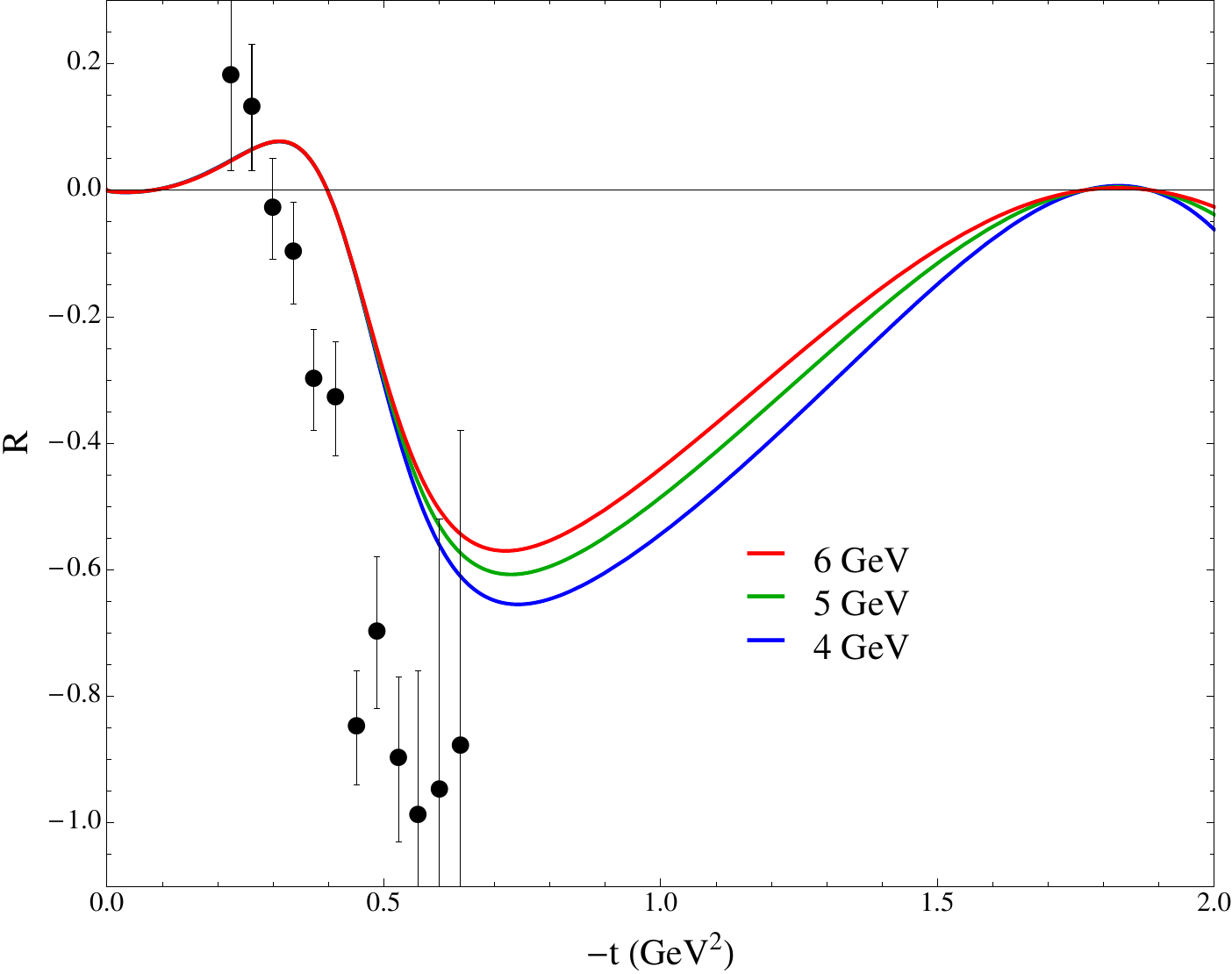}
	\caption {(Color online). Target asymmetry at $E_\gamma=4$ GeV from \cite{Booth:1972qp,Bienlein:1973pt} (left) and recoil asymmetry from \cite{Deutsch:1973tg} (right).}
        \label{fig:dataTA}
\end{figure*}

The target~\cite{Booth:1972qp,Bienlein:1973pt} and recoil asymmetries~\cite{Deutsch:1973tg} are compared to the data  in Fig.~\ref{fig:dataTA}. 
The two target asymmetry measurements~\cite{Booth:1972qp,Bienlein:1973pt} were both performed at $E_\gamma=4$ GeV. The two data sets are not completely compatible, the data from Ref.~\cite{Booth:1972qp} being somehow below the data from Ref.~\cite{Bienlein:1973pt} at small momentum transfers. However, the data presented on Fig.~3 in the original publication~\cite{Booth:1972qp} present a minimum $T\ge -0.7$ where the data from the same publication taken from the Durham data, displayed on the left panel in Fig.~\ref{fig:dataTA}, extend to lower values $T\ge -0.8$. We do not have an explanation for this discrepancy. 
Concerning the recoil asymmetry, the data from Ref.~\cite{Deutsch:1973tg} are given in the energy range from $E_\gamma=4.1$ GeV to  $E_\gamma=6.3$ GeV and we compare with model predictions at  $E_\gamma=4$, 5 and 6 GeV. The target and recoil asymmetries are compatible with each other indicating that amplitude $F_4$ is small {\it cf.} Eq.~\eqref{eq:CS}. 
Although, in the recoil asymmetry,  there is a structure around $t\sim-0.5$ GeV$^2$ absent in the target asymmetry. As emphasized by Berger and Fox \cite{Berger:1971sn}, polarization observables provide crucial information on amplitudes and can discriminate between different high energy models.  In our case, accurate measurements on both $R$ and $T$ polarization observables would improve our knowledge on the poorly known $F_4$ amplitude. Since we have set  $F_4=0$ the target and recoil asymmetries in Eq.~\eqref{eq:CS} are equal and proportional to $\text{Im}\, F_3 F_1^*$. Non-vanishing imaginary part requires at least two amplitudes are present, {\it e.g.} a pole and a cut. Since the vector exchange produces a zero at the nonsense point $\alpha=0$ our model predicts that both asymmetries change sign at $t=-0.4$ GeV$^2$,  which is qualitatively consistent with the data. 
  
Beside the inclusion of the $F_4$ amplitudes, there is another possibility to improve the agreement with the recoil and target polarization data. It might happen that one of the vector poles, $\omega$ or $\rho$, does not have a wrong signature zero in one or both residues  ($F_1$ and/or $F_3$). In this case the contribution of that particular exchange would not vanish at $\alpha_V=0$. In Sec.~\ref{sec:model} we have explained that the wrong signature point $\alpha=0$ is a non-sense point and 
used this observation to justify a zero of the amplitude.  This theoretical expectation should in principle be verified, for example via FESR. For example, a wrong signature zero is present in the $\rho$ pole amplitude in pion-nucleon scattering and consistent with FESR~\cite{Mathieu:2015gxa}. This zero implies a dip in the differential cross section in $\pi^- p \to \pi^0 n$. The dip in $\gamma p \to \pi^0 p$ was therefore also assumed to come from the wrong signature zero.  However the theoretical statement used in Sec.~\ref{sec:model} to justify the wrong signature zero equally applies to $\eta$ photoproduction. Since the neutral pion and eta photoproduction share the same $t-$channel exchanges, the fact that a dip is not observed in the differential cross section in $\gamma p \to \eta p$ ~\cite{Braunschweig:1970jb} indicates that corrections to the pole approximation are stronger in photoproduction and could fill in the non sense zeros. 
  Our model for the $t-$dependence of the poles and cuts, like any other models, has to be checked eventually against analyticity constraints, {\it i.e.} FESR. We hope that FESR in photoproduction, providing constrains on the residues, will shed more light and possibly solve the issues in polarization observables at high energies.

It has been argued in Ref.~\cite{Guidal:1997hy} that the interference between $\rho$ and $\omega$ exchanges properly describes the target and recoil asymmetry. The authors of Ref.~ \cite{Guidal:1997hy} used a rotating phase for the $\rho$ pole, {\it i.e.} $R\propto \exp(-i\pi\alpha)$ instead of the signature factor, {\it i.e.} $R_\rho\propto 1-\exp(-i\pi\alpha)$. The rotating phase emerges as a result 
 of adding two degenerated Regge poles with opposite signature. 
This happens, for example  in charged pion photoproduction where the $a_2$ and $\rho$ exchanges compensate mutually.  More explicitly
\begin{equation}
R_{a_2} - R_\rho \propto (1+e^{-i\pi\alpha}) - (1-e^{-i\pi\alpha})  = 2 e^{-i\pi\alpha}.
\end{equation}
However charge conjugation in neutral pion photoproduction prevents the exchange of the degenerate partners of the $\rho$ and $\omega$ poles (the $a_2$ and $f_2$ poles). Therefore the use of rotating (or constant) phase in neutral pion photoproduction is not justifiable on first principles. 
The only possibility for producing an interference between $\omega$ and $\rho$ pole in polarization observables is a non degeneracy between their trajectories. This could possibly be investigated when more data on neutron target is provided. 

We add for completeness that a weak-Regge-cut model failed to reproduce the target asymmetry as well~\cite{Goldstein:1973xn}

\begin{figure}[htb!]
	\includegraphics[width=\linewidth]{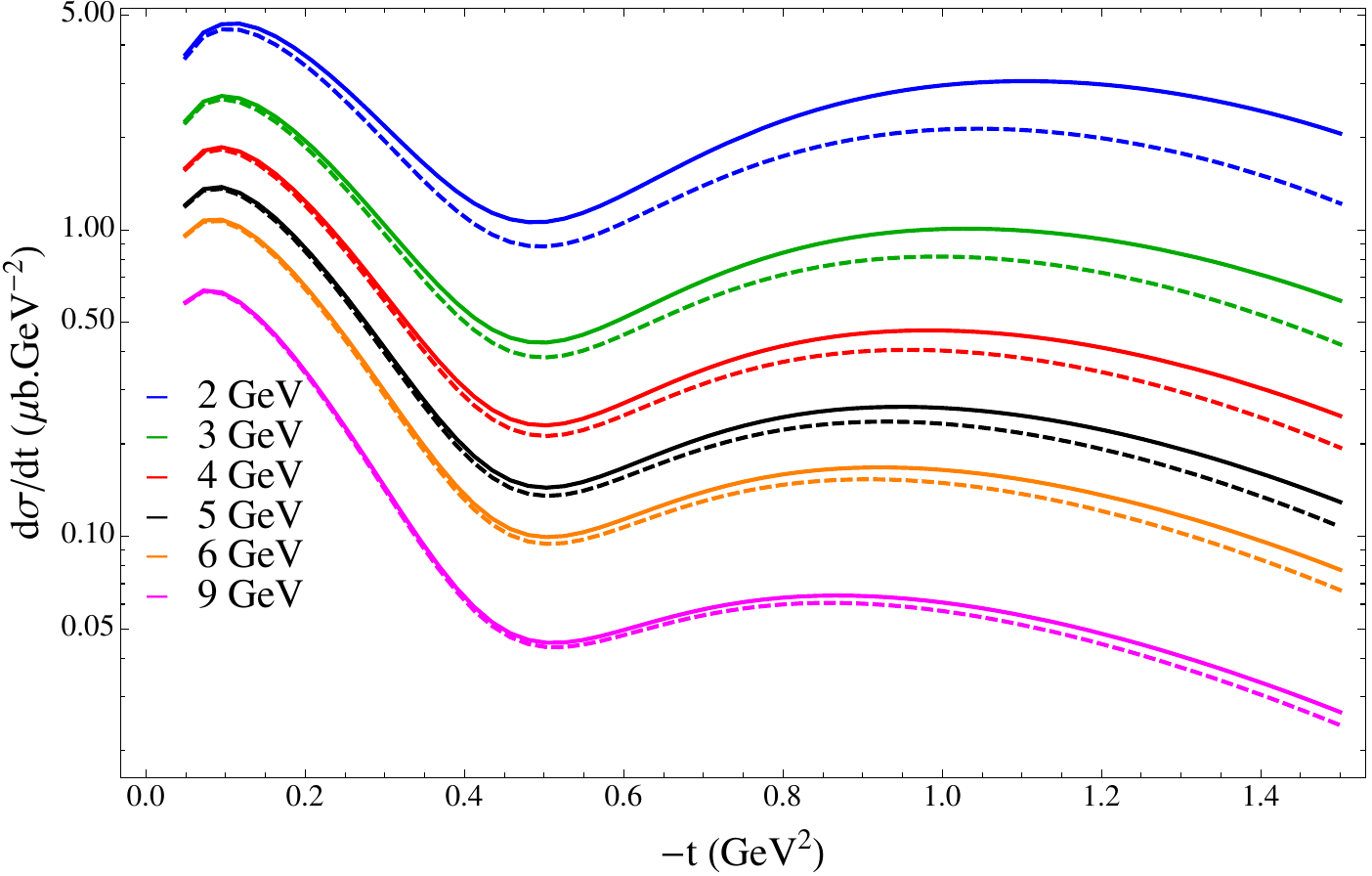}
    \caption {(Color online). Comparison of differential cross section between the high energy limit Eq.~\eqref{eq:CS} (solid lines) and the full expression Eq.~\eqref{eq:CSfull} (dashed lines) for a wide range in incident beam energies.}
    \label{fig:compare}
\end{figure}

We conclude this section with comparison, shown in Fig.~\ref{fig:compare},  of the differential cross section computed using   the exact expression and the high-energy approximation given by  Eq.~\eqref{eq:CSfull} and Eq.~\eqref{eq:CS}, respectively. We observe that the high-energy limit is a good approximation even at the lowest energies considered in the fits. The discrepancy increases as $|t|$ grows, for example,  at $E_\gamma=6$ GeV, and $t=-1$ GeV$^2$ the difference is approximately $12\%$. 

\section{Predictions}\label{sec:pred} 

\begin{figure*}[htb!]
	\includegraphics[width=0.49\linewidth]{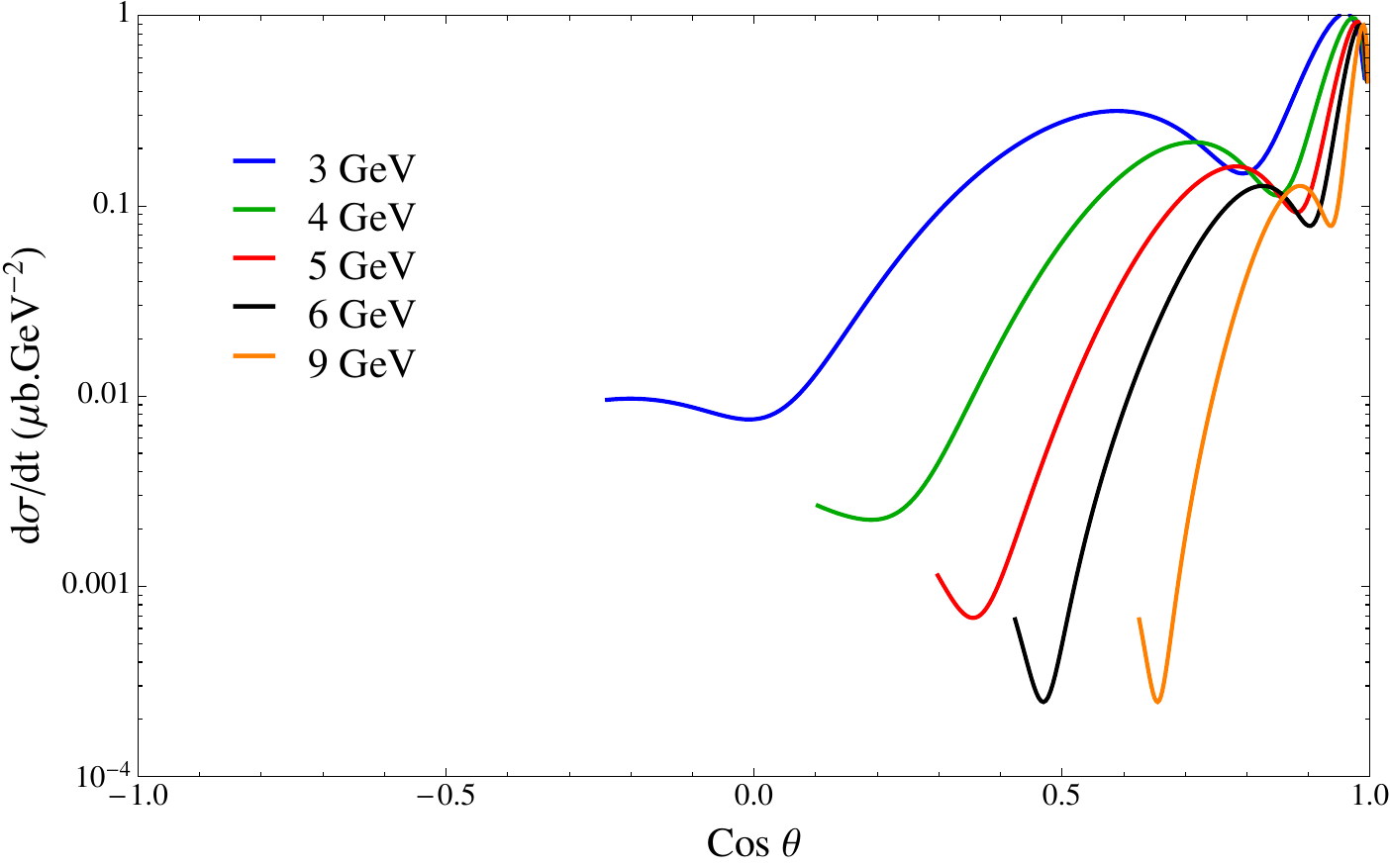}
	\includegraphics[width=0.49\linewidth]{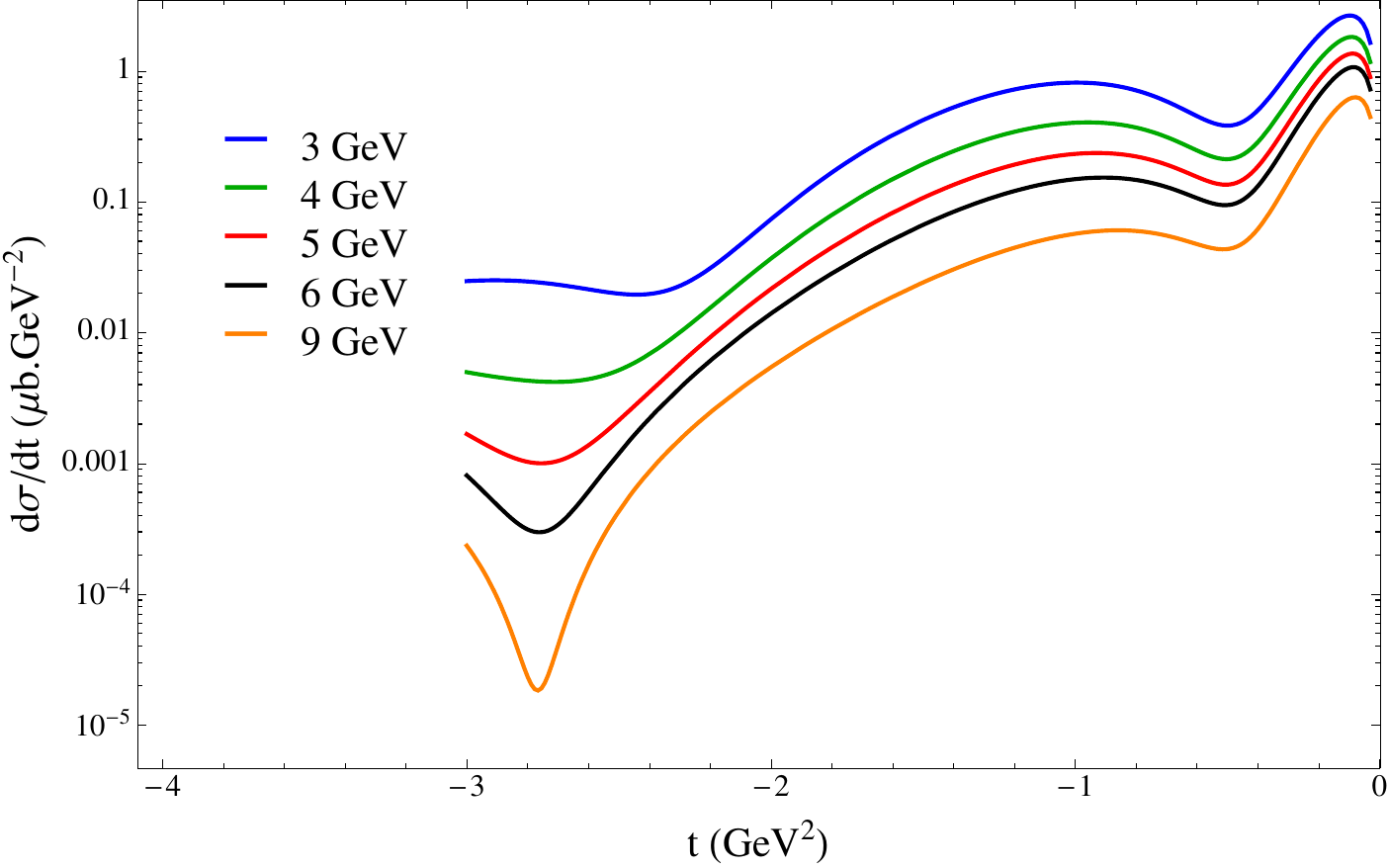}
        \caption {Model prediction (solid lines) for the differential cross section at CLAS and GlueX energies. The dashed lines represent the extrapolation outside the region ($E_\gamma\ge 4$ GeV, $\cos\theta_s\ge0.6$). }
        \label{fig:clas}
\end{figure*}

We give predictions of the model at various beam energies in Fig.~\ref{fig:clas}. The energy range  $E_\gamma=3-6$ GeV corresponds to the recent CLAS measurement~\cite{CLAS}. We also give a prediction for higher energy $E_\gamma = 9$ GeV relevant for the forthcoming measurements at GlueX. We show the differential cross section both as a function of momentum transfer and the $s$-channel  scattering angle in the range $|t| < 3 $ GeV$^2$. 
In the description of the model we repeatedly emphasized the role of zeros in  Regge pole residues,  {\it cf.} Eq.~\eqref{eq:Regge}. These correspond to nonsense points, with $\alpha$ equal to non-positive integer. The first zero appears in the vector trajectory, at $\alpha_V = 0$, and is well established empirically, seen as a dip in the differential cross section at $t\sim-0.5$ GeV$^2$.

At larger $|t|$, zeros corresponding to lower integer values of Regge trajectories should become visible.  For the vector trajectory, $\alpha_V(t)=-2$ corresponds to $t\sim -1.9$ GeV$^2$ and 
 zero of the cut, $\alpha_c(t)=0$ arises at $t\sim-2.75$~GeV$^2$. Since the cut dominates over the pole for large momentum transfers we expect the differential cross section to dip in this region of momentum transfer. Indeed for energies above $E_\gamma=4$ GeV, the dip appears, {\it  cf.} Fig.~\ref{fig:clas}, around $t\sim-2.75$~GeV$^2$.  There are however only a few data points at large $|t|$ to make a detailed comparison wit the model. The second minimum in  the differential cross  section can therefore be used to discriminate between various models for the subleading Regge contributions interfering with the dominant vector pole. In the energy range $E_\gamma=4-5.5$ GeV, the second dip should arise at angle $\theta\sim 60^\circ-80^\circ$, and should be visible in the CLAS data~\cite{CLAS}. 


\section{Conclusion}\label{sec:concl}

We investigated photoproduction of neutral pions for energies above $s$-channel resonance 
 region. Our amplitudes  include the leading Regge poles in the $t$-channel and describe well the differential cross section for beam energies above 4 GeV and for small scattering angles $\cos\theta_s\ge0.6$. 
The first dip seen at $t\sim-0.5$ GeV$^2$, is characteristic to the vector Regge pole and   seems to persist down to $E_\gamma=3$ GeV although the quantitative analysis is not possible given the quality of the data in the  intermediate energy region $3-4$ GeV. Below $E_\gamma = 3$ GeV nucleon resonances become visible and were the focus of most of the recent efforts in pion photo production~\cite{Drechsel:2007if,Workman:2011vb,Anisovich:2012ct,Kamano:2013iva}. These studies could benefit from the results of the higher energy data analysis, for example by implementing finite-$t$ or finite energy sum rule constraints on the resonance models. The resulting baryon spectrum from the analysis in Refs~\cite{Anisovich:2012ct,Kamano:2013iva} is compared in Ref.~\cite{Pennington:2014rra}. A common agreement is found for the lowest nucleon excitations, but both spectra strongly disagree in the center-of-mass energies $1.7-2.0$~GeV. We hope that our amplitudes will provide significant insights in the baryon spectrum to solve the remaining ambiguities in this topic. 
The sum rules for pion photoproduction were studied in the past \cite{Walker:1968xu,Odorico:1976wn,Armenian:1974xd,Collins:1980dv}, however, only now precise partial waves for the low energy data are available.

In Sec.~\ref{sec:data} we analyzed the energy dependence of the beam asymmetry for the reaction $\gamma p \to \pi^0 p$ and the differential cross section for the reaction  $\gamma n \to \pi^0 n$. Based on the expected trajectories for the $\omega, \rho$ and the $b$ poles and the dominance of these Regge poles, we concluded that the beam asymmetry should approach one as the energy increases and the differential cross section for $\gamma n \to \pi^0 n$ should have the same energy dependence as for the reaction $\gamma p \to \pi^0 p$. Both are only qualitatively reproduced, however, the data  for these observables at high energies is rather sparse and we hope that more precise data, including the forthcoming data from CLAS, and the implementation of analyticity constrains will help clarifying these issues. 

All the material, including data and software are available in an interactive from online~\cite{website}.  We invite the interested readers to contact the authors.



\acknowledgments 
We thank R. Workman for his help in developing the website. We acknowledge I. Strakovsky and M. Amaryan for discussions concerning forthcoming CLAS data. 
This material is based upon work supported in part by the U.S. Department of Energy, Office of Science, Office of Nuclear Physics under contract DE-AC05-06OR23177. This work was also supported in part by the U.S. Department of Energy under Grant No. DE-FG0287ER40365, National Science Foundation under Grant PHY-1415459. 

\appendix
\section{$t-$channel helicity amplitudes} \label{app:t}
In this Appendix we compute the combinations of scalar amplitudes with good quantum numbers of the $t$-channel  $\gamma(k) \pi^0(-q)\to \bar N(-p_2) N(p_4)$. 
 The Mandelstam variables $s=(k+p_2)^2$, $t=(k-q)^2$, $u=(k-p_4)^2$  are related through $s+t+u=2M^2+\mu^2$. In the $t-$channel, the physical  domain of the Mandelstam variable is $(t \ge 4M^2, s\le 0)$. 

We start by decomposing the $t-$channel helicity amplitudes in the tensor basis~\cite{Chew:1957tf}. 

\begin{align} \label{eq:A1}
A^t_{\lambda_4\lambda_2,\lambda_1}(s,t) = \bar u_{\lambda_4}(p_4) \sum_{i=1}^4 A_i(s,t) M_i v_{\lambda_2}(-p_2).
\end{align}
where $A_i$ are scalar functions and 
\begin{subequations} \label{eq:defM}
\begin{align}
M_1 &= \frac{1}{2}\gamma_5 \gamma_\mu\gamma_\nu F^{\mu\nu}, \\ 
M_2 &=  2\gamma_5 q_\mu p_\nu F^{\mu\nu},\\
M_3 &=  \gamma_5 \gamma_\mu q_\nu   F^{\mu\nu},\\
M_4 &=\frac{i}{2} \varepsilon_{\alpha\beta\mu\nu} \gamma^\alpha q^\beta F^{\mu\nu},
\end{align}
\end{subequations} 
with $p=(p_2+p_4)/2$. The tensors $F^{\mu\nu}= \epsilon^\mu(k,\lambda_1) k^\nu -  k^\mu  \epsilon^\nu(k,\lambda_1)$ satisfy gauge invariance  by construction. 

In the $t$-channel center of mass frame, 
\begin{align} \nonumber
k^\mu & = (k_t,0,0,k_t), & p_2^\mu &= (-E^t_N,-p_t\sin\theta_t,0,-p_t\cos\theta_t), \\
q^\mu & = (-E^t_\pi,0,0,k_t), & p_4^\mu &= (+E^t_N,-p_t\sin\theta_t,0,-p_t\cos\theta_t).
\end{align}
The first component is the energy and the metric is diag$(+,-,-,-)$. $E^t_N$ and $E^t_\pi$ are the nucleon and pion energies. The scattering angle in the $t-$channel is $\theta_t$. $p_t$ is the momentum of the nuclei in their rest frame. In this frame, $k_t$ is the photon and pion momenta. The masses of the pion and the nucleon are respectively $\mu$ and $M$. The kinematical quantities are
\begin{subequations} \label{eq:kint}
\begin{align}
E^t_\pi &= (t+\mu^2)/2 \sqrt{t} & E^t_N &= \sqrt{t}/2\\
p_t &=\sqrt{t/4-M^2} & k_t&=(t-\mu^2)/2\sqrt{t} \\
\cos\theta_t & = \frac{s-u}{4 k_t p_t} & \sin\theta_t & = \frac{\sqrt{\phi/t}}{2 k_t p_t},
\end{align}
\end{subequations}
with $\phi= stu  - \mu^2M^2 (\mu^2-t) - t M^4 >0 $. 
In the $s$-channel physical region we use the convention of Trueman and Wick Ref.~\cite{Trueman:1964zzb}, and evaluate the square root with the prescription $s\to s+i\epsilon$ and $t\to t-i\epsilon$. It is useful to remember that, in the $s-$channel, $E^t_N$, $k_t$ and $p_t$ are purely imaginary with a negative imaginary part, $\cos\theta_t$ is real and negative and $\sin\theta_t$ is imaginary with a positive imaginary part. 

For the spinors, we use the Dirac representation. The $\gamma-$matrices are, with $\sigma_i$ the Pauli matrices,
\begin{align}
\gamma^0 &= \begin{pmatrix}  1_2 & 0 \\ 0 & -1_2\end{pmatrix}, &
\gamma^i &= \begin{pmatrix}  0 & \sigma^i \\ -\sigma^i & 0\end{pmatrix}, & 
\gamma^5 &= \begin{pmatrix}  0 & 1_2 \\ 1_2 & 0\end{pmatrix}.
\end{align}
For the evaluation of Eq.~\eqref{eq:A1}, the spinors are (with the lower-script $\pm\equiv\pm\frac{1}{2}$)
\begin{align} \nonumber
v_+(-p_2) &= \begin{pmatrix} -\sqrt{E^t_N-M}\ \chi_{2}(\theta_t) \\ \phantom{+}\sqrt{E^t_N+M}\ \chi_{2}(\theta_t)  \end{pmatrix} ,
\\ 
v_-(-p_2) &= \begin{pmatrix} -\sqrt{E^t_N-M}\ \chi_{1}(\theta_t)  \\ -\sqrt{E^t_N+M}\ \chi_{1}(\theta_t) \end{pmatrix} ,
\\  \nonumber
\bar u_+(p_4) &= \begin{pmatrix} \sqrt{E^t_N+M}\ \chi_{2}^\dag(\theta_t)  & -\sqrt{E^t_N-M}\ \chi^\dag_{2}(\theta_t) \end{pmatrix},  \\  \nonumber
\bar u_-(p_4) &= \begin{pmatrix} \sqrt{E^t_N+M}\ \chi_{1}^\dag(\theta_t)  & \phantom{+}\sqrt{E^t_N-M}\ \chi_{1}^{\dag}(\theta_t)  \end{pmatrix},
\end{align}
with 
\begin{align} 
\chi_1(\theta)   &=  \begin{pmatrix}  \cos\theta/2 \\  \sin\theta/2\end{pmatrix} , &
\chi_2(\theta)   &=  \begin{pmatrix}  -\sin\theta/2 \\  \phantom{+}\cos\theta/2\end{pmatrix} ,
\end{align}
and the polarization tensor for the photon is $\epsilon^\mu(k,\pm 1) = (0, \mp1 , -i , 0)/\sqrt{2}$.

With these definition the $t-$channel amplitudes can be expressed in terms of the scalar amplitudes and the kinematical quantities, 
\begin{align} \nonumber
A^t_{++,1} &=\sqrt{2} k_t \frac{\sin\theta_t}{2} \left[ \sqrt{t} \left(A_1-2M A_4\right) - 2 p_t \left(A_1+tA_2\right) \right] \\  \nonumber
A^t_{--,1} &= \sqrt{2} k_t \frac{\sin\theta_t}{2} \left[ \sqrt{t} \left(A_1-2M A_4\right) + 2 p_t \left(A_1+tA_2\right) \right] \\  \nonumber
A^t_{+-,1} &= \sqrt{2} k_t \sin^2\frac{\theta_t}{2}\left[\phantom{+}2 p_t  \sqrt{t} A_3 -\left(2M A_1-t A_4\right) \right]\\
A^t_{-+,1} &= \sqrt{2} k_t \cos^2 \frac{\theta_t}{2}\left[-2 p_t  \sqrt{t} A_3 -\left(2M A_1-t A_4\right)\right]
 \label{eq:TAmpl}
\end{align}
The amplitudes with negative photon helicity are obtained from the relations
\begin{align}
A^t_{\pm\pm,-1} &=  A^t_{\mp\mp,1} ,& A^t_{\pm\mp,-1} &=  -A^t_{\mp\pm,1}.
\end{align}
In Eq.~\eqref{eq:TAmpl},  the invariant amplitudes $A_i$ contain dynamical singularities. Some of the kinematical singularities in $t$ are explicitly extracted (and arise in $p_t, k_t$ and $\sqrt{t}$). All kinematical singularities in the variable $s$ are encoded in the trigonometric functions. They arise from the spin of the external particles and are independent of the exchanged particle. They can be extracted easily from the partial wave decomposition in the $t-$channel
\begin{align}\label{pwatspin2}
A^t_{\lambda_4\lambda_2,\lambda_1} (s,t) & = \sum_{J=1}^\infty (2J+1) T^J_{\lambda'\lambda}(t)\, d_{\lambda'\lambda}^J(z_t).
\end{align}
with $ \lambda=\lambda_1-\lambda_3=\lambda_1$, $\lambda'=\lambda_2-\lambda_4$ and $z_t = \cos\theta_t$. The Wigner rotation function 
 $d_{\lambda'\lambda}^J(z_t)=\xi_{\lambda'\lambda}(z_t)P_{\lambda'\lambda}^J(z_t)$, where $P^J$ is a polynomial, has indeed a cut in the $s$ variable. They originate from the half-angle factor
\be
\xi_{\lambda'\lambda}(z_t) = \left(\frac{1-z_t}{2}\right)^{\frac{1}{2} |\lambda'-\lambda| } \left(\frac{1+z_t}{2}\right)^{\frac{1}{2} |\lambda'+\lambda| }.	
\ee
The partial waves $T^J_{\lambda'\lambda}$ have a well defined spin $J$ but are not eigenstate of  parity. The good parity combinations are  $T^J_{\lambda'\lambda}\pm T^J_{\lambda'-\lambda}$. In order to form parity conserving amplitudes free of kinematical singularities, one first needs to remove the half-angle factor with the definition
\begin{align} 
A^t_{\lambda_4\lambda_2,\lambda_1} = \left[\cos \frac{\theta_t}{2}\right]^{|\lambda+\lambda'|}\left[\sin \frac{\theta_t}{2}\right]^{|\lambda-\lambda'|} \widehat T_{\lambda\lambda'}.
\end{align}
Then $t-$channel parity conserving helicity amplitudes (PCHAs) are given by 
\cite{collins, spearman}
\begin{align}
\widehat T^\eta_{\lambda'\lambda} = \frac{1}{\sqrt{2}} \left(\widehat T_{\lambda'\lambda} +\eta(-1)^{\lambda'}\ \widehat T_{\lambda'-\lambda} \right).
\end{align}
$\eta$ is called the naturality as it corresponds to $P(-)^J$ for an exchanged particle of parity $P$ and spin $J$ in the $t-$channel. 
Using the expressions~\eqref{eq:TAmpl}, we obtain the relations between the PCHAs and the invariant amplitudes
\begin{subequations}
\begin{align}
\widehat T^+_{01} &=  - 2 k_t \sqrt{t} (-A_1+2M A_4) \\
\widehat T^-_{01} &= -4 p_t k_t(A_1+t A_2) \\
\widehat T^+_{11} &=  -2 k_t (2MA_1-t A_4) \\
\widehat T^-_{11} &=  \phantom{+}4 p_t k_t \sqrt{t} A_3
\end{align}
\end{subequations}
The quantum numbers of the PCHAs are best computed using the standard non relativistic state $|J,M,L,S\rangle$, {\it cf.} Appendix of Ref.~\cite{Childers:1963}. Using Eq.~(B5) from ~\cite{Jacob:1959at}, we can express the 2-nucleon state $|J,M;\lambda_4,\lambda_2\rangle$ as
\begin{subequations}
\begin{align} \nonumber
\sqrt{2} | J,0;\pm\pm\rangle =& \pm| J,0, J,0 \rangle \\  \nonumber
& + \left(\frac{J}{2J+1}\right)^{\frac{1}{2}}| J,0, J-1,1\rangle \\  
&- \left(\frac{J+1}{2J+1}\right)^{\frac{1}{2}}| J,0,J+1, 1\rangle, \\  \nonumber
\sqrt{2} | J,\pm1;\pm\mp\rangle =& \mp| J,\pm1, J,1 \rangle \\  \nonumber
& + \left(\frac{J+1}{2J+1}\right)^{\frac{1}{2}}| J,\pm1, J-1, 1\rangle \\  
&+ \left(\frac{J}{2J+1}\right)^{\frac{1}{2}}| J,\pm1, J+1, 1\rangle.
\end{align}
\end{subequations}
Since $|J,M,L,S\rangle$ have parity $(-1)^{L+1}$ and charge conjugation $(-1)^{L+S}$, the PCHAs are invariant under $CP$. In the above decomposition, only $| J,0, J,0 \rangle$ has $CP=-1$.  Introducing standard combinations of invariant amplitudes, we find that their quantum numbers are 
\begin{subequations}\label{eq:Finv2}
\begin{align}
F_1 &= -A_1+2MA_4, & \eta&=+1, & CP&=+1,\\
F_2&= A_1+t A_2 , & \eta&=-1, & CP&=-1,\\
F_3&= 2M A_1-t A_4, & \eta&=+1, & CP&=+1,\\
F_4 &=A_3,  & \eta&=-1, & CP&=+1.
\end{align}
\end{subequations}

\section{$s-$channel helicity amplitudes and observables} \label{app:s}
We are interested in the observables for the photoproduction of a neutral pion at the leading order in the center of mass energy squared. The observables are functions of the $s-$channel amplitudes, defined by
\begin{align}\label{eq:As}
A^s_{\mu_4,\mu_2\mu_1} &= \bar u_{\mu_4}(p_4) \sum_{i=1}^4 A_i M_i\  u_{\mu_2}(p_2).
\end{align}
In Walker's notation \cite{Walker:1968xu} we have ($W=\sqrt{s}$)
\begin{subequations}
\begin{align}
A^s_{+,+1}  & = (8\pi W) H_4, &  
A^s_{+,-1}  & = (8\pi W) H_3, \\
A^s_{-,-1}  & = (8\pi W) H_1, & 
A^s_{-,+1}  & = (8\pi W) H_2.
\end{align}
\end{subequations}
$H_1$ and $H_4$ are single spin flip amplitudes, $H_2$ is the non-flip amplitude and $H_3$ is the double flip amplitude.

We use SAID/MAID conventions for the observables~\cite{Workman:2011hi}
\begin{subequations}\label{eq:CS-App}
\begin{align}
\frac{d\sigma}{dt} & = \frac{\pi}{k_s^2}\ \frac{1}{2} \left(|H_1|^2 + |H_2|^2 + |H_3|^2 + |H_4|^2 \right), \\
\Sigma\frac{d\sigma}{dt} & = \frac{\pi}{k_s^2}\ \text{Re}\left(H_1 H_4^* - H_2 H_3^*\right), \\
T\frac{d\sigma}{dt} & = \frac{\pi}{k_s^2}\ \text{Im}\left(H_1 H_2^* + H_3 H_4^*\right),  \\
R\frac{d\sigma}{dt} & = \frac{\pi}{k_s^2}\ \text{Im}\left(H_3 H_1^* + H_4 H_2^*\right).
\end{align}
\end{subequations}
$k_s=(s-M^2)/2\sqrt{s}=M E_\gamma/\sqrt{s}$ is the photon momentum in the $s$-channel 
 center of mass frame.

The $s-$channel amplitudes can be evaluated using an explicit representation for the spinors as we did in the previous section. In the $s-$channel region, the four vector are
\begin{align} \nonumber
k^\mu & = (k_s,0,0,\phantom{+}k_s), & q^\mu &= (E_\pi^s,\phantom{+}q_s\sin\theta_s,0,\phantom{+}q_s\cos\theta_s), \\
p_2^\mu & = (E_2^s,0,0,-k_s), & p_4^\mu &= (E_4^s,-q_s\sin\theta_t,0,-q_s\cos\theta_t).
\end{align}
The kinematical quantities are
\begin{align} \nonumber
k_s &= (s-M^2)/2 \sqrt{s}, & E_\pi^s &= (s-M^2+\mu^2) /2\sqrt{s},\\ \nonumber
E_2^s &=(s+M^2)/2 \sqrt{s}, & E_4^s &= (s+M^2-\mu^2) /2\sqrt{s},\\
 \cos\theta_s & = \frac{t-u + \Delta/s}{4 k_s q_s},  &\sin\theta_s & = \frac{\sqrt{\phi/s}}{2 k_s q_s},
\end{align}
with $\Delta = M^2(M^2-\mu^2)$ and 
\begin{align}
q_s & = \left[ (s-(M+\mu)^2)(s-(M-\mu)^2) \right]^{\frac{1}{2}}/2\sqrt{s}.
\end{align}
For the evaluation of Eq.~\eqref{eq:As}, the spinors are 
\begin{align} \nonumber
u_+(p_2) &= \begin{pmatrix}\phantom{+} \sqrt{E_2^s+M}\ \chi_2(0) \\ \phantom{+} \sqrt{E_2^s-M}\ \chi_2(0)\end{pmatrix} ,
\\
u_-(p_2) &= \begin{pmatrix}- \sqrt{E_2^s+M}\ \chi_1(0) \\ \phantom{+} \sqrt{E_2^2-M}\ \chi_1(0) \end{pmatrix} ,
\\ \nonumber
\bar u_+(p_4) &= \begin{pmatrix} \sqrt{E_4^2+M}\ \chi_2^\dag(\theta_s) & -\sqrt{E_4^s-M}\ \chi_2^\dag(\theta_s) \end{pmatrix},  \\ \nonumber
\bar u_-(p_4) &= \begin{pmatrix} \sqrt{E_4^s+M}\ \chi_1^\dag(\theta_s) &\phantom{+} \sqrt{E_4^s-M}\ \chi_1^\dag (\theta_s)\end{pmatrix},
\end{align}
and the polarization tensor for the photon is $\epsilon^\mu(k,\pm 1) = (0, \mp1 , -i , 0)/\sqrt{2}$. 

The expression for the $s-$channel helicity amplitudes in terms of the CGLN invariant amplitudes $A_i$ are quite lengthy. Their expressions are most conveniently expressed as function of other scalar amplitudes, the CGLN ${\cal F}_i$. We obtain
\begin{subequations}
\begin{align} 
H_1 &=\frac{-1}{\sqrt{2}} \sin\theta_s  \cos\frac{\theta_s}{2} \left( {\cal F}_3 + {\cal F}_4\right) \\  
H_3 &= \frac{1}{\sqrt{2}} \sin\theta_s  \sin\frac{\theta_s}{2} \left( {\cal F}_3 - {\cal F}_4\right) \\ 
H_2 &= \sqrt{2}\cos\frac{\theta_s}{2}   \left( {\cal F}_2 - {\cal F}_1\right)  + H_3\\  
H_4 &= \sqrt{2}\sin\frac{\theta_s}{2}   \left( {\cal F}_2 + {\cal F}_1\right)  - H_1
\end{align}
\end{subequations}
where
\begin{subequations}
\begin{align} \nonumber
{\cal F}_1  = \frac{Z_2^+Z_4^+}{8\pi W} \Big[& -(M-W)A_1 + \frac{\mu^2-t}{2}(A_3-A_4) \\
& + (M-W)^2 A_4  \Big] \\ \nonumber
{\cal F}_2  = \frac{Z_2^-Z_4^-}{8\pi W} \Big[& - (M+W)A_1 + \frac{\mu^2-t}{2}(A_3-A_4) \\
& + (M+W)^2 A_4  \Big] \\ \nonumber
{\cal F}_3  = q_s \frac{Z_2^-Z_4^+}{8\pi W} &(M+W) \Big[ - (M-W)A_2 \\  &+ (A_3-A_4)\Big] \\ \nonumber
{\cal F}_4  = q_s \frac{Z_2^+Z_4^-}{8\pi W} &(M-W) \Big[ - (M+W)A_2 \\ &+ (A_3-A_4)\Big] 
\end{align}
\end{subequations}
It worth noting that the amplitudes ${\cal F}_i$ have kinematical singularities coming from the factors
\begin{align}
Z_{2,4}^\pm = \sqrt{E_{2,4} \pm M}.
\end{align}

Another instructive method to obtain the $s-$channel amplitudes, given in \cite{Trueman:1964zzb}, is to express them in terms of the $t-$channel using crossing relations.  Of course, parity conserving combinations rotate independently
\begin{align}\label{eq:crossing}
\begin{pmatrix} A^s_{+,+1} \pm A^s_{-,-1} \\  A^s_{+,-1} \mp A^s_{-,+1}  \end{pmatrix} & = -i R\left(\frac{\chi_2\mp\chi_4}{2} \right)
\begin{pmatrix} A^t_{++,1} \pm A^t_{--,1} \\  A^t_{+-,1} \mp A^t_{-+,1}  \end{pmatrix}.
\end{align}
The rotation matrix is 
\begin{align}
R(\chi) = \begin{pmatrix} \phantom{+}\cos \chi& \sin\chi \\ 
-\sin \chi& \cos  \chi \end{pmatrix},
\end{align}
and the crossing angles are
\begin{align} \nonumber
\cos\chi_2 &= -\frac{-st+M^2(\mu^2-t)}{\lambda^{\frac{1}{2}} (s,0,M^2)\lambda^{\frac{1}{2}} (t,0,\mu^2)}\approx -\sqrt{\frac{-t}{4M^2-t}}, \\
\cos\chi_4 &= \frac{(-t)(s-\mu^2) - M^2(t+2\mu^2)}{\lambda^{\frac{1}{2}} (s,\mu^2,M^2)\lambda^{\frac{1}{2}} (t,M^2,M^2)} \approx \sqrt{\frac{-t}{4M^2-t}}.
\end{align}
The symbols $\approx$ stands for the leading order in $s$ and the triangle function is $\lambda(a,b,c)=a^2+b^2+c^2-2(ab+bc+ca)$. Since the crossing matrix \eqref{eq:crossing} is a rotation (up to a global phase\footnote{This phase is present in the original calculation in Ref.~\cite{Trueman:1964zzb} but was overlooked in subsequent publications. See for instance Ref.~\cite{collins}. }), it preserves the norm. The differential cross section can be then computed directly with the $t-$channel amplitudes without performing the rotation. The result is
\begin{align}\nonumber
 \frac{d\sigma}{dt} =  \frac{1}{64\pi} &\frac{|k_t|^2}{4 M^2 E_\gamma^2} \bigg[
\phantom{+} 2|\sin\theta_t|^2 \left( |2 p_t F_2|^2 - t |F_1|^2 \right) \\ \nonumber
& +  (1-\cos\theta_t)^2 \left| F_3+2\sqrt{t} p_t F_4\right|^2 \\ 
& +  (1+\cos\theta_t)^2 \left| F_3-2\sqrt{t} p_t F_4\right|^2  \bigg]. 
\end{align}
In the evaluation of the differential cross section with $t-$channel amplitudes, one has to remember that $\sin\theta_t$ is a complex number in the physical region of the process $\gamma p \to \pi^0 p$. 

Single polarization observables $\Sigma,T,R$ are best evaluated in the high energy limit. The error made by using this approximation are compensated by calculating the ratio of quadratic forms of amplitudes, cf. Eq.~\eqref{eq:CS-App}. Keeping only the dominant term in $s$, the crossing relations take a simple form
\begin{align} \nonumber
\frac{1}{\sqrt{2}}\begin{pmatrix} H_4+H_1\\  H_3-H_2  \end{pmatrix} & \approx   \frac{W/8\pi}{4M^2-t}\begin{pmatrix} \phantom{+}2M&  \sqrt{-t} \\  -\sqrt{-t} & 2M \end{pmatrix}
\begin{pmatrix} \sqrt{-t} F_1 \\  \phantom{ \sqrt{-t} } F_3 \end{pmatrix},\\
\frac{1}{\sqrt{2}}\begin{pmatrix} H_4-H_1\\  H_3+H_2  \end{pmatrix} & \approx  
\frac{W}{8\pi}\begin{pmatrix} 0&  1 \\ -1 & 0 \end{pmatrix}
\begin{pmatrix} \phantom{ \sqrt{-t} } F_2 \\  \sqrt{-t}F_4 \end{pmatrix}.
 \label{eq:rott2s}
\end{align}
The apparent pole at $t=4M^2$ in the first equation above is spurious and disappears when the scalar amplitudes $A_i$ are substituted. We indeed obtain
\begin{align} 
\frac{1}{\sqrt{2}}\begin{pmatrix} H_4+H_1\\  H_3-H_2  \end{pmatrix} & \approx  \frac{W}{8\pi}
\begin{pmatrix} \sqrt{-t} A_4 \\  \phantom{ \sqrt{-t} } A_1 \end{pmatrix}.
\end{align}
The relations \eqref{eq:rott2s} were derived in Ref.~\cite{Barker:1974vm}. We corrected a sign mistake in the rotation matrix Eq.~(2.3) in Ref.~\cite{Barker:1974vm}. This mistake propagated through the observables. The reader could easily check that the pole $t=4M^2$ does not cancel as it should in the observables Eq.~(A.10)-(A.13) in Ref.~\cite{Barker:1974vm}. 
The correct expressions at leading order in the energy squared are
\begin{subequations}\label{eq:ObsLeading2}
\begin{align}
 \frac{d\sigma}{dt} &\approx  \frac{1}{32\pi} \left[\frac{|F_3|^2- t|F_1|^2}{4M^2- t} + |F_2|^2 - t |F_4|^2 \right] \\
\Sigma \frac{d\sigma}{dt} &\approx  \frac{1}{32\pi} \left[\frac{|F_3|^2- t|F_1|^2}{4M^2- t} - |F_2|^2 + t |F_4|^2 \right] \\
T \frac{d\sigma}{dt} &\approx  \frac{1}{16\pi} \sqrt{- t}\ \text{Im}\,\left[\frac{F_3F_1^*}{4M^2- t} + F_4F_2^* \right]\\
R \frac{d\sigma}{dt} &\approx  \frac{1}{16\pi} \sqrt{- t}\ \text{Im}\,\left[\frac{F_3F_1^*}{4M^2- t} - F_4F_2^* \right]
\end{align}
\end{subequations}


\end{document}